\documentclass[12pt,a4paper]{article}
\usepackage[T1]{fontenc}

\usepackage{geometry, tabularx}
\normalsize
\usepackage{amsmath,amssymb,amsthm,bm,mathtools,xspace,booktabs,url}
\usepackage{graphicx,etoolbox}
\usepackage{ dsfont }
\usepackage{bbold}
\usepackage{tikz}
\usepackage{ mathrsfs }
\usepackage[shortlabels]{enumitem}
\usepackage{xcolor}
\usepackage{color}

\newcommand{\dd}{\mathrm{d}}
\makeatletter
\global\let\tikz@ensure@dollar@catcode=\relax
\makeatother
\setlist{
  listparindent=\parindent,
  parsep=0pt,
}
\usepackage{amssymb}

\usepackage{hyperref}
\usepackage{cleveref}

\AtBeginEnvironment{thebibliography}{
  \small
  \setlength\itemsep{0.75em plus 0.5em minus 0.75em}%
}

\usepackage{caption}
\usepackage[raggedright]{titlesec}

\numberwithin{equation}{section}
\usepackage{amsfonts}

\theoremstyle{plain} 
\newtheorem{theorem}{Theorem}[section]
\newtheorem{Lemma}{Lemma}[section]
\newtheorem{Proposition}{Proposition}[section]

\newtheorem{definition}{Definition}[section]
\newtheorem*{theorem*}{Theorem}

\theoremstyle{definition} 
\newtheorem{Example}{Example}[section]
\newtheorem{Remark}{Remark}[section]

\allowdisplaybreaks

\usepackage[above,below,section]{placeins}

\usepackage[all]{onlyamsmath}

\usepackage{multirow}

\usepackage{xcolor}
\usepackage{color}

\usepackage{mathrsfs}
\usepackage{pdfpages}

\definecolor{darkmagenta}{rgb}{0.5,0,0.5}
\definecolor{darkgreen}{rgb}{0,0.6,0}
\definecolor{darkblue}{rgb}{0,0,0.6}
\definecolor{darkred}{rgb}{0.8,0,0}
\definecolor{mellow}{rgb}{.847, 0.72, 0.525}

\begin{document}

\title{Multivariate matrix-exponential affine mixtures and their applications in risk theory}

\author{Eric C.K. Cheung\footnote{School of Risk and Actuarial Studies, UNSW Business School, University of New South Wales, Sydney, NSW 2052, Australia. \texttt{eric.cheung@unsw.edu.au}}, Oscar Peralta\footnote{School of Mathematical Sciences, The University of Adelaide, Adelaide, SA 5005, Australia. \texttt{oscar.peraltagutierrez@adelaide.edu.au} (corresponding author)} \thanks{Department of Actuarial Science, Faculty of Business and Economics, University of Lausanne, CH-1015 Lausanne, Switzerland.}, Jae-Kyung Woo\footnote{School of Risk and Actuarial Studies, UNSW Business School, University of New South Wales, Sydney, NSW 2052, Australia. \texttt{j.k.woo@unsw.edu.au}}
}
\date{}

\maketitle
\vspace{-.3in}
\begin{abstract}
In this paper, a class of multivariate matrix-exponential affine mixtures with matrix-exponential marginals is proposed. The class is shown to possess various attractive properties such as closure under size-biased Esscher transform, order statistics, residual lifetime and higher order equilibrium distributions. This allows for explicit calculations of various actuarial quantities of interest. The results are applied in a wide range of actuarial problems including multivariate risk measures, aggregate loss, large claims reinsurance, weighted premium calculations and risk capital allocation. Furthermore, a multiplicative background risk model with dependent risks is considered and its capital allocation rules are provided as well. We finalize by discussing a calibration scheme based on complete data and potential avenues of research.

\end{abstract}\vspace{-.2in}

\vspace{-.2in}

\section{Introduction}\vspace{-.2in}

Modelling of multivariate risks or losses has always been an important topic in actuarial research. Suppose that an insurance portfolio has $M$ dependent risks collected in the vector $\bm{X}=(X_1,\dots,X_M)$. Classical actuarial problems include risk aggregation (e.g. \cite{Cossette2013FGMErlang, Cossette2015BivExp}) as well as the development and analysis of risk measures based on individual loss $X_j$ or the total loss $S=\sum_{j=1}^MX_j$ (see e.g. \cite{ArtznerMF1999, goovaerts2010RM} for the univariate case and \cite{Cousin2014MultiCTE, Cossette2016MultiTVaR} for the multivariate case). Treating the risk measure of the aggregate risk $S$ as the risk capital that the insurer should possess in order to remain solvent with high probability, it is of the insurer's interest to decide how much capital to allocate to each individual risk. This leads to the research problem of capital allocation, where the allocation rules should satisfy some desirable properties (e.g. \cite{Denault2001RAC, Dhaene2003RAC, Tsanakas2009CAconvex, Dhaene2012RAC, FurmanZitikis2008weightCA}), further considered in the case a multiplicative factor randomly scales each individual risk \cite{zhu2012asymptotic,asimit2013evaluating,merz2014demand}. Risk measures are also related to premium principles \cite{Furman2006PremElliptical}, and in particular weighted premium calculations \cite{FurmanZitikis2008weightPC}. The determination of (re)insurance premiums such as excess-of-loss, stop-loss and large claims (re)insurance for a portfolio is also a non-trivial task (see e.g. \cite{AlbrecherBeirlantTeugels2017ReinBook} for a review).

When solving the afore-mentioned actuarial problems, the risks $\bm{X}=(X_1,\dots,X_M)$ are often assumed to follow specific multivariate distributions so that explicit formulas for the quantities of interest can be derived. In particular, closure properties under convolutions, residual lifetime distributions, size-biased Esscher transforms and order statistics can be helpful for various applications because calculations for the transformed variables can be easily performed with a change of parameters. From a practical point of view, it is also important to be able to fit the multivariate distributions to data using algorithms that can be easily implemented. Examples of multivariate distributions that have enjoyed success in some or all of these aspects include mixed Erlang \cite{lee2012modeling,willmot2015some} (which is also related to joining exponential, Erlang or mixed Erlang marginal distributions via the Farlie-Gumbel-Morgenstern copula or Sarmanov's family \cite{Cossette2013FGMErlang, Cossette2015BivExp, Ratovomirija2017Sarmanov}), gamma \cite{Furman2005RCAgamma, Furman2021MultiMixGamma}, Pareto \cite{Asimit2010MultiPareto, Sarabia2016weightCA}, elliptical \cite{Valdez2003CAelliptical, Furman2006PremElliptical, landsman2016multivariate, landsman2018multivariate}, and phase-type \cite{CaiLi2005MPHCTE}.

In this paper, we shall make use of matrix-exponential distributions, which are less explored in the actuarial literature, and propose a class of multivariate matrix-exponential affine mixtures for use in actuarial science. The class of univariate matrix-exponential distributions \cite{bladt1992renewal,bean2008characterization,bladt2017matrix} serves as a natural generalization of the class of phase-type distributions \cite{bladt2005review}. However, despite the similar representation of the densities of the two classes, analysis of matrix-exponential distributions mostly relies on analytic methods. This is in contrast to phase-type distributions where probabilistic interpretation is common, often in connection to Markov chains and level-crossing arguments. There exist different classes of multivariate distributions with matrix-exponential or phase-type marginals in the literature with varying degrees of generality and tractability. For instance, in \cite{assaf1984multivariate} the authors defined a class of multivariate phase-type distributions whose marginals correspond to the hitting times of different sets in a finite-state space Markov jump process, with a generalization defined in \cite{kulkarni1989new} using a system of rewards for the occupation times in the underlying process. A further generalization to multivariate matrix-exponential distributions was proposed in \cite{bladt2010multivariate}, where the authors provided a characterization for the class of multivariate distributions with rational multivariate Laplace transform. Of these families, the one in \cite{assaf1984multivariate} is the only one for which an explicit form of its multivariate density functions is known, which limits the applicability of those found in \cite{kulkarni1989new} and \cite{bladt2010multivariate} drastically. This prompts the need to define tractable and robust families of multivariate distributions whose marginals have an explicit matrix-exponential-like structure which facilitates calculations and actuarial applications.



This paper is organized as follows. Section \ref{sec:ME} provides a review of univariate matrix-exponential distributions, which includes some basic properties and useful techniques that will be used throughout the paper. Simple proofs of some existing and new results will also be given. In Section \ref{sec:MMEm}, we introduce our proposed multivariate matrix-exponential affine mixtures which possess matrix-exponential marginals. Closure properties with regard to size-biased Esscher transform, order statistics as well as conditional, residual lifetime and higher order equilibrium distributions are presented along with related results on cross moments and rank correlations. Section \ref{sec:Applications} demonstrates applications of the results in a wide range of actuarial problems including multivariate risk measures, aggregate loss, risk capital allocation, large claims reinsurance and weighted premium calculations. We also include a study of the capital allocation rules for a multiplicative background risk model in the case where the risks, before being scaled by a random multiplicative factor, exhibit dependence. Finally, in Section \ref{sec:calibration} we conclude the paper by providing a calibration scheme based on the use of Bernstein copulas. Some of its advantages and disadvantages will be discussed along with further avenues of research.
\vspace{-.2in}

\section{Matrix-exponential distributions}\label{sec:ME}\vspace{-.2in}

Let $X$ be an absolutely continuous random variable on $[0,\infty)$ which has a density function $f$ of the form
\begin{equation}\label{eq:densityME1}f(x)=\bm{\alpha}e^{\bm{T}x}\bm{t}
\end{equation}
for $x\ge 0$, where $\bm{\alpha}$ is a $p$-dimensional real row vector, $\bm{T}$ a $p\times p$ real square matrix, and $\bm{t}$ a $p$-dimensional real column vector, for some $p\in \mathds{N}_+$ with $\mathds{N}_+$ the set of positive integers. In such a case, we say that $X$ follows a \emph{matrix-exponential} ($\mathrm{ME}$) distribution, denoted by $X\sim\mathrm{ME}(\bm{\alpha},\bm{T},\bm{t})$, where $(\bm{\alpha},\bm{T},\bm{t})$ is the triple associated to the distribution. Let us denote by $\mathrm{ME}$ the family of probability density functions of the form (\ref{eq:densityME1}), which we call the class of $\mathrm{ME}$ density functions. For later use, the cumulative distribution function associated to the density $f$ is denoted by $F(x)=\int_0^x f(y)\dd y$ for $x\ge 0$ and the survival function is $\overline{F}=1-F$. When multiple ME densities are concerned in the context, the triple of the $j$-th density $f_j\in\mathrm{ME}$ will be denoted by $(\bm{\alpha}_j,\bm{T}_j,\bm{t}_j)$, and the corresponding cumulative distribution function and survival function are $F_j$ and $\overline{F}_j$, respectively. A few important facts about the $\mathrm{ME}$ class are summarized below (see \cite{bladt1992renewal} or \cite[Chapter 4]{bladt2017matrix} for details).\vspace{-.2in}
\begin{itemize}
\item Different triples can yield the same $\mathrm{ME}$ density function. In other words, the parametrization of a given $\mathrm{ME}$ distribution is not unique.
\item The class $\mathrm{ME}$ coincides with the class of probability density functions with support on $[0,\infty)$ which have a rational Laplace transform. In particular, if the numerator and the denominator of the Laplace transform of a given $\mathrm{ME}$-distributed random variable are relatively prime and the order of the denominator equals $p$, then the $\mathrm{ME}$ representation is said to be minimal.
\item The matrix $\bm{T}$ associated to an $\mathrm{ME}(\bm{\alpha},\bm{T},\bm{t})$ distribution can be chosen in such a way that it has a strictly negative real eigenvalue, say $\kappa<0$, such that it dominates all other (possibly complex) eigenvalues $\{\sigma_i\}$, i.e. $\kappa \ge \Re(\sigma_i)$.
\item The cumulative distribution function associated to (\ref{eq:densityME1}) is of the form 
\begin{equation}\label{eq:distME}
F(x)=1-\bm{\alpha}e^{\bm{T}x}\bm{l}
\end{equation}
for $x\ge 0$, where $\bm{l}=(-\bm{T})^{-1}\bm{t}$.
\item Let $\{\sigma_i\}$ be the collection of (possibly complex) eigenvalues of $\bm{T}$, and let $m_i$ be the multiplicity associated to the eigenvalue $\sigma_i$. Then, the density function (\ref{eq:densityME1}) can be written as a linear combination, with possibly negative coefficients, of 
\begin{equation}\label{eq:spanME1}\{x^{k}e^{\sigma_i x} x: i; 1\le k\le m_i\}.\end{equation}
Moreover, any density function contained in the linear span of (\ref{eq:spanME1}) corresponds to an $\mathrm{ME}$ distribution.
\end{itemize}\vspace{-.2in}

A common subclass of $\mathrm{ME}$ used in applied probability is that of phase-type distributions, the case in which $\bm{\alpha}$ corresponds to a probability row vector, $\bm{T}$ a sub-intensity matrix, and $\bm{t}=-\bm{T}\bm{1}$ where $\bm{1}$ denotes a column vector of ones. In this context, $X$ can be understood as the termination time of a Markov jump process driven by $\bm{T}$ with initial distribution $\bm{\alpha}$. Note, however, that the class of phase-type distributions is strictly smaller than $\mathrm{ME}$. A canonical example \cite[Example  4.5.21]{bladt2017matrix} of a distribution which is $\mathrm{ME}$ but not phase-type is given by the density function $f(x) = (2/3)e^{-x}(1 + \cos(x))$: 
one of its $\mathrm{ME}$ representations is described by the triple
\[\bm{\alpha}=(1,0,0),\quad \bm{T} = \begin{pmatrix} -1& -1 & 2/3\\ 1& -1& -2/3\\ 0& 0 & -1\end{pmatrix}, \quad \bm{t}=\begin{pmatrix}4/3 \\ 2/3 \\ 1\end{pmatrix},\]
while a phase-type representation is inexistent \cite{o1990characterization}. In recent years, the use of strict $\mathrm{ME}$ distributions (that is, non-phase-type) has attracted attention in the literature since its convergence to the unit mass measure is faster (with respect to its dimension) than that achieved by the classic Erlangization technique \cite{horvath2020numerical}.

It is well known that the class of $\mathrm{ME}$ distributions inherits some properties of phase-type distributions, though these properties are usually proved by analytic means rather than by the probabilistic ones commonly used for the phase-type case (e.g. see \cite{bladt2005review}). Below we present an example involving the convolution of $\mathrm{ME}$-distributed random variables. Although a proof of this can be found in e.g. \cite[Theorem 4.4.2]{bladt2017matrix} using Laplace transforms, in what follows we provide an alternative simple proof using an identity related to integrals of matrix quantities.
\begin{Proposition}\label{prop:convolutionME1}
(\textbf{Closure of $\mathrm{ME}$ class under convolutions}) For $1\le j\le n$, let $X_j\sim\mathrm{ME}(\bm{\alpha}_j,\bm{T}_j,\bm{t}_j)$ be independent, and let $\bm{k}=\{k_1,\dots,k_m\}\subset\{1,\dots,n\}$ with $1\le m\le n$. Then, $\sum_{j\in\bm{k}}X_{j}$ is $\mathrm{ME}$-distributed with parameters $(\bm{\alpha}_{\bm{k}},\bm{T}_{\bm{k}}, \bm{t}_{\bm{k}})$ where
\[
\bm{\alpha}_{\bm{k}} = (\bm{\alpha}_{k_1},\bm{0},\bm{0},\dots,\bm{0}),\quad \bm{T}_{\bm{k}}=\begin{pmatrix}\bm{T}_{k_1} & \bm{t}_{k_1} \bm{\alpha}_{k_2}&&\\&\bm{T}_{k_2} & \bm{t}_{k_2} \bm{\alpha}_{k_3}&\\&&\ddots & \\&&&\bm{T}_{k_m}\end{pmatrix},\quad \bm{t}_{\bm{k}}=\begin{pmatrix}\bm{0}\\\bm{0}\\\vdots\\\bm{t}_{k_m}\end{pmatrix},
\]
with $\bm{0}$ being a matrix of zeroes of appropriate dimension.
\end{Proposition}\vspace{-.2in}
\begin{proof}
Fix $x\ge 0$ and suppose that $m\ge 2$ (the case $m=1$ being trivial). Following \cite[Theorem 1]{van1978computing} with regard to integrals of matrix exponentials, we get
\begin{align*}
\bm{\alpha}_{\bm{k}}&e^{\bm{T}_{\bm{k}} x}\bm{t}_{\bm{k}}= \bm{\alpha}_{k_1}\begin{pmatrix}\bm{I}& 
\bm{0}&\cdots&\bm{0}\end{pmatrix}e^{\bm{T}_{\bm{k}} x}\begin{pmatrix}\bm{0}\\ 
\bm{0}\\\vdots\\\bm{I}\end{pmatrix}\bm{t}_{k_m}\\
& = \int_0^x\int_0^{y_1} \cdots \int_0^{y_{m-2}} \bm{\alpha}_{k_1}e^{\bm{T}_{k_1}(x-y_1)} \bm{t}_{k_1} \bm{\alpha}_{k_2} e^{\bm{T}_{k_2}(y_1-y_2)} \bm{t}_{k_2} \bm{\alpha}_{k_3}\cdots e^{\bm{T}_{k_m}y_{m-1}}\bm{t}_{k_m}\dd y_{m-1}\cdots\dd y_2\dd y_1\\
& = \int_0^x\int_0^{y_1} \cdots \int_0^{y_{m-2}} \left(\bm{\alpha}_{k_1}e^{\bm{T}_{k_1}(x-y_1)} \bm{t}_{k_1}\right) \left(\bm{\alpha}_{k_2} e^{\bm{T}_{k_2}(y_1-y_2)} \bm{t}_{k_2}\right) \cdots \left(\bm{\alpha}_{k_m}e^{\bm{T}_{k_m}y_{m-1}}\bm{t}_{k_m}\right)\dd y_{m-1}\cdots\dd y_2\dd y_1\\
& = \int_0^x\int_0^{y_1} \cdots \int_0^{y_{m-2}} f_{k_1}(x-y_1) f_{k_2}(y_1-y_2)\cdots f_{k_m}(y_{m-1})\dd y_{m-1}\cdots\dd y_2\dd y_1\\
& = f_{k_1}\ast f_{k_2}\ast\cdots\ast f_{k_m}(x),
\end{align*}
where $\bm{I}$ denotes an identity matrix of appropriate dimension, $f_j$ is an $\mathrm{ME}(\bm{\alpha}_j,\bm{T}_j,\bm{t}_j)$ density, and `$\ast$' the convolution operation. Thus, the result follows.
\end{proof}\vspace{-.2in}

In the following, we consider two classes of mixture distributions which contain $\mathrm{ME}$. First we define the class of \emph{matrix-exponential mixtures} ($\mathrm{MEm}$) by
\begin{equation*}
\mathrm{MEm} = \left\{\sum_{j=1}^L c_j f_j : L\in\mathds{N}_+; f_j\in\mathrm{ME}; c_j\ge 0; \sum_{j=1}^L c_j=1\right\},
\end{equation*}
which contains density functions on $[0,\infty)$ that represent proper mixtures of $\mathrm{ME}$ densities (which must be non-negative). Further define the seemingly more general class of \emph{matrix-exponential affine mixtures} ($\mathrm{MEam}$) by
\begin{equation}\label{eq:MEam1} \mathrm{MEam} = \left\{\sum_{j=1}^L c_j f_j : L\in\mathds{N}_+; f_j\in\mathrm{ME}; c_j\in\mathds{R}; \sum_{j=1}^L c_j=1; \sum_{j=1}^L c_j f_j \ge0\right\},\end{equation}
which allows (some of) the ``mixing weights'' $c_j$'s to be possibly negative as long as $\sum_{j=1}^L c_j f_j$ is non-negative.  In the actuarial science literature, mixtures with negative weights are often referred to as combinations \cite{dufresne2007CombExp}. The following Proposition \ref{prop:eqclass} shows that both $\mathrm{MEm}$ and $\mathrm{MEam}$ are equivalent to $\mathrm{ME}$. This allows for a mixture representation of the densities that will help us define a new class of tractable multivariate distributions in Section \ref{sec:MMEm}, which is the main goal of this paper. 
\begin{Proposition}\label{prop:eqclass}
(\textbf{Closure of $\mathrm{ME}$ class under (affine) mixtures}) We have the equivalence $\mathrm{ME}=\mathrm{MEm}=\mathrm{MEam}$.
\end{Proposition}\vspace{-.2in}
\begin{proof}
Clearly $\mathrm{ME}\subseteq\mathrm{MEm}\subseteq\mathrm{MEam}$, so all that is left to prove is that $\mathrm{MEam}\subseteq\mathrm{ME}$. Pick an element of $\mathrm{MEam}$ defined in (\ref{eq:MEam1}) where each $f_j$ has parameters $(\bm{\alpha}_j,\bm{T}_j,\bm{t}_j)$. Then, for all $x\ge 0$ we have
\[
\sum_{j=1}^L c_j f_j(x)=\left(c_1 \bm{\alpha}_1,\dots, c_L \bm{\alpha}_L\right)\exp\left(\begin{pmatrix}\bm{T}_1&&\\&\ddots&\\&&\bm{T}_L\end{pmatrix}x\right)\begin{pmatrix}\bm{t}_1\\ \vdots\\ \bm{t}_L\end{pmatrix},
\]
with the right-hand side corresponding to an element of $\mathrm{ME}$. 
\end{proof}\vspace{-.2in}

Even though the classes $\mathrm{ME}$, $\mathrm{MEm}$ and $\mathrm{MEam}$ are the same, it may be convenient to express densities in a given particular form depending on the context. For example, although an $\mathrm{ME}$ representation of the order statistics of $\mathrm{ME}$-distributed independent random variables can be found in \cite[Section 4.4.2]{bladt2017matrix}, 
it is considerably high dimensional. By considering $\mathrm{MEam}$ representations instead, we can avoid the computational cost of calculating the matrix exponential function of a very large matrix in exchange of computing a finite sum of lower dimensional matrix exponential functions.  Proposition \ref{prop:order1} below presents the results for the order statistics where an $\mathrm{MEam}$ representation is simpler than the $\mathrm{ME}$ one.
\begin{Proposition}\label{prop:order1}
(\textbf{Closure of $\mathrm{ME}$ under order statistics}) For $1\le j\le n$, let $X_j\sim \mathrm{ME}(\bm{\alpha}_{j},\bm{T}_j,\bm{t}_j)$ be independent, and let $X_{k:n}$ denote the $k$-th order statistic of the collection $\{X_1, \dots, X_n\}$ (in the sense that $X_{1:n}\le X_{2:n}\le \dots \le X_{n:n}$). Then, the density function $f_{k:n}$ of $X_{k:n}$ is an element of $\mathrm{MEam}$. 
\end{Proposition}\vspace{-.2in}
\begin{proof}
Let $\mathscr{P}_n$ be the set of all permutations of $\{1,\dots, n\}$ and let $\mathscr{B}_\ell=\{0,1\}^\ell$ for $0\le \ell\le n$. If $\bm{r}\in\mathscr{P}_n$, we denote its entries by $(r_1,\dots, r_n)$; similarly, we denote the entries of $\bm{q}\in\mathscr{B}_\ell$ by $(q_1,\dots, q_\ell)$. By the general formula for order statistics \cite[Equation (8)]{balakrishnan2007permanents}, for $x\ge 0$ and $1\le k\le n$,
\begin{align}
f_{k:n}(x)&=\frac{1}{(k-1)! (n-k)!} \sum_{\bm{r}\in\mathscr{P}_n}F_{r_1}(x)\cdots F_{r_{k-1}}(x) f_{r_k}(x)\overline{F}_{r_{k+1}}(x)\cdots \overline{F}_{r_{n}}(x)\nonumber\\
&=\frac{1}{(k-1)! (n-k)!} \sum_{\substack{\bm{r}\in\mathscr{P}_n\\\bm{q}\in \mathscr{B}_{k-1}}}(-1)^{z(\bm{q})}\left(\overline{F}_{r_1}(x)\right)^{q_1}\cdots \left(\overline{F}_{r_{k-1}}(x)\right)^{q_{k-1}} f_{r_k}(x)\overline{F}_{r_{k+1}}(x)\cdots \overline{F}_{r_{n}}(x),
\label{eq:orderstats1}
\end{align}
where 
$z(\bm{q})=\sum_j q_j$. Thus, $f_{k:n}\in\mathrm{MEam}$ will follow once we prove that each summand on the right-hand side of (\ref{eq:orderstats1}) is proportional to some density which belongs to $\mathrm{ME}$. 

Fix $1\le k\le n$, $\bm{r}\in\mathscr{P}_n$ and $\bm{q}\in \mathscr{B}_{k-1}$, and let $\mathcal{S}=\{r_j: 1\le j\le k-1; q_{j}=1\}$. The set $\mathcal{S}$ contains exactly $z=z(\bm{q})$ elements, which we call $s_1,\dots,s_z$. Denote by $\otimes$ and $\oplus$ the Kronecker product and sum \cite[A.4]{bladt2017matrix}, respectively. Using that $\overline{F}_j(x)=\bm{\alpha}_je^{\bm{T}_j x}\bm{l}_j$ where $\bm{l}_j=(-\bm{T}_j)^{-1}\bm{t}_j$, we obtain
\begin{align}
& \left(\overline{F}_{r_1}(x)\right)^{q_1}\cdots \left(\overline{F}_{r_{k-1}}(x)\right)^{q_{k-1}} f_{r_k}(x)\overline{F}_{r_{k+1}}(x)\cdots \overline{F}_{r_{n}}(x)\label{eq:orderaux1}\\
& \quad = \overline{F}_{s_1}(x)\cdots \overline{F}_{s_{z}}(x) f_{r_k}(x) \overline{F}_{r_{k+1}}(x)\cdots \overline{F}_{r_n}(x)\nonumber\\
& \quad = \left(\bm{\alpha}_{s_1}e^{\bm{T}_{s_1}x}\bm{l}_{s_1}\right)\cdots\left(\bm{\alpha}_{s_{z}}e^{\bm{T}_{s_{z}}x}\bm{l}_{s_{z}}\right)\left(\bm{\alpha}_{r_{k}}e^{\bm{T}_{r_{k}}x}\bm{t}_{r_{k}}\right)\left(\bm{\alpha}_{r_{k+1}}e^{\bm{T}_{r_{k+1}}x}\bm{l}_{r_{k+1}}\right)\cdots\left(\bm{\alpha}_{r_{n}}e^{\bm{T}_{r_{n}}x}\bm{l}_{r_{n}}\right)\nonumber\\
&\quad = \left(\bm{\alpha}_{s_{1}}\otimes\cdots\otimes\bm{\alpha}_{s_{z}}\otimes \bm{\alpha}_{r_{k}}\otimes \bm{\alpha}_{r_{k+1}}\otimes\cdots\otimes \bm{\alpha}_{r_{n}}\right)\nonumber\\
&\qquad\quad \times \left(e^{\bm{T}_{s_{1}}x}\otimes\cdots\otimes e^{\bm{T}_{s_{z}}x} \otimes e^{\bm{T}_{r_{k}}x}\otimes e^{\bm{T}_{r_{k+1}}x}\otimes\cdots\otimes e^{\bm{T}_{r_{n}}x}\right)\nonumber\\
&\qquad\quad \times \left(\bm{l}_{s_{1}}\otimes\cdots\otimes\bm{l}_{s_{z}}\otimes \bm{t}_{r_{k}}\otimes \bm{l}_{r_{k+1}}\otimes\cdots\otimes \bm{l}_{r_{n}}\right)\nonumber\\
&\quad = \left(\bm{\alpha}_{s_{1}}\otimes\cdots\otimes\bm{\alpha}_{s_{z}}\otimes \bm{\alpha}_{r_{k}}\otimes \bm{\alpha}_{r_{k+1}}\otimes\cdots\otimes \bm{\alpha}_{r_{n}}\right)\nonumber\\
&\qquad\quad \times \left(\exp\left(\left(\bm{T}_{s_{1}}\oplus\cdots\oplus\bm{T}_{s_z}\oplus \bm{T}_{r_{k}}\oplus \bm{T}_{r_{k+1}}\oplus\cdots\oplus\bm{T}_{r_{n}}\right)x\right)\right)\nonumber\\
&\qquad\quad \times \left(\bm{l}_{s_{1}}\otimes\cdots\otimes\bm{l}_{s_{z}}\otimes \bm{t}_{r_{k}}\otimes \bm{l}_{r_{k+1}}\otimes\cdots\otimes \bm{l}_{r_{n}}\right),\nonumber
\end{align}
where the second-to-last equality follows from the mixed product rule for Kronecker products \cite[Theorem A.4.2]{bladt2017matrix}, and the last equality from a Kronecker sum exponential identity \cite[Theorem A.4.7]{bladt2017matrix}. That (\ref{eq:orderaux1}) is proportional to a density function follows by noticing that such expression is non-negative and bounded from above by $f_{r_k}$ (and thus integrable).
\end{proof}
\begin{Remark}\label{rem:orderstats2}
Note that if $X_1, \dots , X_n$ follow an identical distribution $\mathrm{ME}(\bm{\alpha},\bm{T},\bm{t})$, then $X_{k:n}$ has density function
\begin{align*}
 f_{k:n}(x)& =\frac{n!}{(k-1)!(n-k)!} \left(F(x)\right)^{k-1} f(x) \left(\overline{F}(x)\right)^{n-k}\\
& =  \frac{n!}{(k-1)!(n-k)!} \sum_{\ell=0}^{k-1} {k-1 \choose \ell} (-1)^{\ell} f(x) \left(\overline{F}(x)\right)^{n-\ell - 1},
\end{align*}
where
\begin{equation*}
f(x) \left(\overline{F}(x)\right)^{n-\ell - 1} = \underbrace{\left(\bm{\alpha}\otimes \bm{\alpha}\otimes\cdots\otimes \bm{\alpha}\right)}_{n-\ell \;\mathrm{blocks}} \exp\big(\underbrace{\left(\bm{T}\oplus\bm{T}\oplus\cdots\oplus\bm{T}\right)}_{n-\ell \;\mathrm{blocks}}x\big)\underbrace{\left(\bm{t}\otimes\bm{l}\otimes\cdots\otimes\bm{l}\right)}_{n-\ell \;\mathrm{blocks}}.
\end{equation*}
\end{Remark}\vspace{-.2in}

\section{Multivariate matrix-exponential affine mixtures}\label{sec:MMEm}\vspace{-.2in}

In this section, we exploit the representation (\ref{eq:MEam1}) to define a class of multivariate distributions with $\mathrm{MEam}$ marginals. Our construction is based on an affine mixture resembling the class multivariate Erlang mixture distributions considered in \cite{lee2012modeling,willmot2015some}. Throughout the rest of the paper, we shall fix $L, M\in\mathds{N}_+$, and for all $j\in\{1,\dots,L\}$, fix $f_j\in \mathrm{ME}$ with triple $(\bm{\alpha}_j,\bm{T}_j,\bm{t}_j)$. An $M$-variate version of our proposed matrix-exponential affine mixture is defined as follows.
\begin{definition}  (\textbf{$\mathrm{MMEam}$})
Define $\mathscr{S}=\{1,\dots, L\}^M$ and let $\{p_{\bm{i}}\}_{\bm{i}\in\mathscr{S}}$ be a collection of real numbers such that $\sum_{\bm{i}\in\mathscr{S}} p_{\bm{i}}=1$, where the entries of $\bm{i}\in\mathscr{S}$ are given by $(i_1,\dots, i_M)$.  With $f_j\in \mathrm{ME}$ for each $j\in\{1,\dots,L\}$, we say that the multivariate density function defined by 
\begin{equation}\label{eq:densityMMEm1}f(x_1,\dots,x_M)=\sum_{\bm{i}\in\mathscr{S}}p_{\bm{i}}f_{i_1}(x_1)\cdots f_{i_M}(x_M)
\end{equation}
for $x_1,\dots,x_M\ge 0$ is a multivariate matrix-exponential affine mixture ($\mathrm{MMEam}$) density function  as long as $f\ge0$. We denote the class of such density functions by $\mathrm{MMEam}$.
\end{definition}\vspace{-.2in}

It is clear from the above definition that if an $M$-variate random vector (where $M\ge 2$) has joint density (\ref{eq:densityMMEm1}), then any $\widetilde{M}$ sub-components (where $1\le \widetilde{M}<M$) of the random vector has joint density that follows an $\widetilde{M}$-variate $\mathrm{MMEam}$ (see the proof of Theorem \ref{th:crossmoments1}).
\begin{Remark}\label{rem:simulation1} If $p_{\bm{i}}\ge 0$ for all $\bm{i}\in\mathscr{S}$, then the random vector $\bm{X}=(X_1,\dots,X_M)$ following the law (\ref{eq:densityMMEm1}) can be easily realized by first drawing a point $\bm{\omega}=(\omega_1,\dots,\omega_M)\in \mathscr{S}$ according to the $M$-variate probability mass function $\{p_{\bm{i}}:\bm{i}\in\mathscr{S}\}$, followed by drawing $Z_k\sim f_{\omega_k}$ independently for all $1\le k\le M$, and finally assigning the values of each $Z_k$ to $X_k$.
\end{Remark}
\begin{Remark}\label{rem:dense} If $p_{\bm{i}}\ge 0$ for all $\bm{i}\in\mathscr{S}$ and  for some $\lambda>0$, $f_j$ is an Erlang($j,\lambda$) density (which belongs to $\mathrm{ME}$) for all $j\in\{1,\dots,L\}$, then (\ref{eq:densityMMEm1}) reduces to the density of a multivariate finite Erlang mixture. Moving from Erlang to $\mathrm{ME}$ density provides added flexibility to modelling, which may be reflected in a smaller number of terms to achieve the same or even better fit. Moreover, while the Erlang($j,\lambda$) distribution has the smallest coefficient of variation among phase-type distributions of order $j$ \cite{david1987least}, even lower coefficient of variations can be found within the $\mathrm{ME}$ class \cite{horvath2020numerical}.
\end{Remark}\vspace{-.2in}

In the rest of this section we investigate some closure properties and related results of the $\mathrm{MMEam}$ class.  For later use, we denote by $\mathds{N}_0$ the set of non-negative integers.\vspace{-.2in}

\subsection{Conditional density and cross moments}\label{sec:ConPDF}\vspace{-.2in}
Suppose that the random vector $(X_1,\dots, X_M)$ follows an $M$-variate density function $f$ with support on $[0,\infty)^M$. Fix $1\le k\le M$ and $x_{k+1}, \dots , x_{M}\ge 0$. Let $f^\mathrm{con}$ denote the joint density of $(X_1,\dots,X_k)$ conditional on $(X_{k+1},\dots, X_M) = (x_{k+1},\dots , x_M)$, which is given by
\begin{equation}\label{eq:ConDensityGen}
f^\mathrm{con}(x_1,\dots , x_k \mid x_{k+1},\dots,x_M) = \frac{f(x_1,\dots,x_M)}{\int_0^\infty \cdots \int_0^\infty f(y_1,\dots,y_k,x_{k+1},\dots,x_M)\dd y_1\cdots\dd y_k}
\end{equation}
for $x_1, \dots , x_k\ge 0$. In the following, we show that if $f\in\mathrm{MMEam}$ then $f^\mathrm{con}$ also belongs to $\mathrm{MMEam}$ and the resulting cross moments can be easily evaluated. When $k=M$, the results simply apply to the (unconditional) full vector $(X_{1},\dots,X_M)$.
\begin{theorem}\label{th:crossmoments1} (\textbf{Conditional density and cross moments of $\mathrm{MMEam}$})
Let $(X_1, \dots, X_M)$ follow the density $f\in\mathrm{MMEam}$ of the form (\ref{eq:densityMMEm1}). Then the conditional density $f^\mathrm{con}$ defined in (\ref{eq:ConDensityGen}) is a $k$-variate $\mathrm{MMEam}$ density given by
\begin{equation}\label{eq:ConDensityMMEam}
f^\mathrm{con}(x_1,\dots , x_k \mid x_{k+1},\dots,x_M) = \sum_{\bm{i}\in\mathscr{S}} q_{\bm{i}}(x_{k+1},\dots,x_M) f_{i_1}(x_1)\cdots f_{i_k}(x_k),
\end{equation}
where 
\begin{equation}\label{eq:aux1ci}q_{\bm{i}}(x_{k+1},\dots,x_M) = \left\{\begin{array}{ccc}\frac{p_{\bm{i}} f_{i_{k+1}}(x_{k+1})\cdots f_{i_{M}}(x_{M})}{\sum_{\bm{h}\in\mathscr{S}} p_{\bm{h}} f_{h_{k+1}}(x_{k+1})\cdots f_{h_{M}}(x_{M})}&\mbox{if}& k<M.\\ p_{\bm{i}} &\mbox{if}& k=M.
\end{array}\right.\end{equation}
In addition, for $r_{1},r_{2},\dots,r_{k}\in\mathds{N}_0$, the cross moments are
\begin{equation}\label{eq:ConCrossMom}
\mathds{E}\left[\left.\prod_{j = 1}^k X_j^{r_j}\;\right| \;X_{k+1}=x_{k+1},\dots, X_M=x_M\right] = \sum_{\bm{i}\in\mathscr{S}} q_{\bm{i}}(x_{k+1},\dots,x_M) \prod_{j = 1}^k r_j!\, \bm{\alpha}_{i_j}(-\bm{T}_{i_j})^{-(r_j+1)}\bm{t}_{i_j}.
\end{equation}
\end{theorem}\vspace{-.2in}
\begin{proof}
We start by considering the joint density of $(X_{k+1}, \dots, X_M)$ for $k<M$. This is represented by the denominator of (\ref{eq:ConDensityGen}), 
which, upon substitution of (\ref{eq:densityMMEm1}), equals
\begin{align}
& \int_0^\infty \cdots \int_0^\infty f(y_1,\dots,y_k,x_{k+1},\dots,x_M)\dd y_1\cdots\dd y_k = \sum_{\bm{h}\in\mathscr{S}} p_{\bm{h}} f_{h_{k+1}}(x_{k+1})\cdots f_{h_{M}}(x_{M})\label{eq:jointM-kdensity0}\\
& \quad = \sum_{(h_{k+1},\dots, h_M)\in\{1,\dots, L\}^{M-k}} \Bigg(\sum_{(h_1,\dots, h_k)\in\{1,\dots, L\}^k} p_{h_1,\dots, h_M}\Bigg) f_{h_{k+1}}(x_{k+1})\cdots f_{h_M}(x_{M}),\label{eq:jointM-kdensity}
\end{align}
where the entries of $\bm{h}\in\mathscr{S}$ are $(h_1,\dots, h_M)$. The above expression clearly follows an $(M-k)$-variate $\mathrm{MMEam}$ with the terms inside the large parenthesis the ``mixing weights''. When $k<M$, via division of (\ref{eq:densityMMEm1}) by (\ref{eq:jointM-kdensity0}), we can express (\ref{eq:ConDensityGen}) as (\ref{eq:ConDensityMMEam}) with $q_{\bm{i}}(x_{k+1},\dots,x_M)$ given by (\ref{eq:aux1ci}). The result on $f^\mathrm{con}$ is trivial when $k=M$. Note that the summation in (\ref{eq:ConDensityMMEam}) can be separated into two parts in the same manner as in (\ref{eq:jointM-kdensity}), and therefore one observes that (\ref{eq:ConDensityMMEam}) is a $k$-variate $\mathrm{MMEam}$ density. Finally,
\begin{align*}
&\mathds{E}\left[\left.\prod_{j = {1}}^k X_j^{r_j}\;\right| \;X_{k+1}=x_{k+1},\dots, X_M=x_M\right]\\
&\quad = \sum_{\bm{i}\in\mathscr{S}} q_{\bm{i}}(x_{k+1},\dots,x_M) \left(\int_{0}^\infty x_1^{r_1} f_{i_1}(x_1)\dd x_1\right)\cdots \left(\int_{0}^\infty x_k^{r_k} f_{i_k}(x_k)\dd x_k\right).
\end{align*}
Since $\int_{0}^\infty x_j^{r_j} f_{i_j}(x_j)\dd x_j = r_j!\, \bm{\alpha}_{i_j}(-\bm{T}_{i_j})^{-(r_j+1)}\bm{t}_{i_j}$ \cite[Theorem 4.1.18]{bladt2017matrix}, the result (\ref{eq:ConCrossMom}) immediately follows.
\end{proof}\vspace{-.3in}

\subsection{Residual lifetime distribution}\label{sec:RL}\vspace{-.2in}
For an $M$-variate density function $f$ with support on $[0,\infty)^M$ that is represented by the random vector $\bm{X}=(X_1,\dots,X_M)\sim f$, fix $\bm{z}=(z_1,\dots,z_M)\in [0,\infty)^M$ and let $f^\mathrm{RL}_{\bm{z}}$ be the multivariate density of $\bm{X}-\bm{z}$ conditional on $\bm{X}>\bm{z}$. We call $f^\mathrm{RL}_{\bm{z}}$ the \emph{residual lifetime density} of $f$. It is straightforward to see that 
\begin{equation}\label{eq:residualGen}
f^\mathrm{RL}_{\bm{z}}(x_1,\dots,x_M) =\frac{f(x_1+z_1,\dots ,x_M+z_M)}{\overline{F}(z_1,\dots, z_M)}
\end{equation}
for $x_1,\dots,x_M\ge 0$, where $\overline{F}(z_1,\dots,z_M)=\int_{z_1}^\infty \dots \int_{z_M}^\infty f(y_1,\dots,y_M)\dd y_1\cdots \dd y_M$ is the joint survival function corresponding to $f$.
\begin{theorem}\label{th:residual1}  (\textbf{Residual lifetime of $\mathrm{MMEam}$}) Suppose that $f\in\mathrm{MMEam}$ is of the form (\ref{eq:densityMMEm1}). Then the residual lifetime density function $f^\mathrm{RL}_{\bm{z}}$ defined in (\ref{eq:residualGen}) is also in the class $\mathrm{MMEam}$ and its explicit form is given by
\begin{equation}\label{eq:residualMMem1}
f^\mathrm{RL}_{\bm{z}}(x_1,\dots,x_M) = \sum_{\bm{i}\in\mathscr{S}} p^\mathrm{RL}_{\bm{z},\bm{i}}\,f^\mathrm{RL}_{z_1,i_1}(x_1) \cdots f^\mathrm{RL}_{z_M,i_M}(x_M),
\end{equation}
where
\begin{equation}\label{eq:residualcom1MMem1}
f^\mathrm{RL}_{z,j}(x)=  \bm{\alpha}^\mathrm{RL}_{z,j}e^{\bm{T}_j x} \bm{t}_j,\quad \bm{\alpha}^\mathrm{RL}_{z,j}=\frac{1}{\bm{\alpha}_je^{\bm{T}_j z} \bm{l}_j} \bm{\alpha}_{j}e^{\bm{T}_j z}
\end{equation}
for $z\ge0$ and $1\le j \le L$, and
\begin{equation}\label{eq:residualcom2MMem1}
p^\mathrm{RL}_{\bm{z},\bm{i}}=\frac{p_{\bm{i}}\left(\bm{\alpha}_{i_1}e^{\bm{T}_{i_1} z_1} \bm{l}_{i_1}\right)\cdots \left(\bm{\alpha}_{i_M}e^{\bm{T}_{i_M} z_M} \bm{l}_{i_M}\right)}{\sum_{\bm{h}\in\mathscr{S}}p_{\bm{h}}\left(\bm{\alpha}_{h_1}e^{\bm{T}_{h_1} z_1} \bm{l}_{h_1}\right)\cdots \left(\bm{\alpha}_{h_M}e^{\bm{T}_{h_M} z_M} \bm{l}_{h_M}\right)}
\end{equation}
under the obvious definition $\bm{l}_j=(-\bm{T}_j)^{-1}\bm{t}_j$.
\end{theorem}\vspace{-.2in}
\begin{proof}
By putting (\ref{eq:densityMMEm1}) into (\ref{eq:residualGen}), we arrive at
\[
f^\mathrm{RL}_{\bm{z}}(x_1,\dots,x_M) = \frac{\sum_{\bm{i}\in\mathscr{S}}p_{\bm{i}}f_{i_1}(x_1+z_1)\cdots f_{i_M}(x_M+z_M)}{\overline{F}(z_1,\dots,z_M)},
\]
which can be rewritten as (\ref{eq:residualMMem1}) by defining
\[
f^\mathrm{RL}_{z,j}(x)=\frac{f_{j}(x+z)}{\overline{F}_{j}(z)},\quad p^\mathrm{RL}_{\bm{z},\bm{i}}= \frac{p_{\bm{i}}\,\overline{F}_{i_1}(z_1)\cdots \overline{F}_{i_M}(z_M)}{\overline{F}(z_1,\dots,z_M)}.
\]
Note that $f^\mathrm{RL}_{z,j}$ represents a univariate residual lifetime density of $f_j$. Since (\ref{eq:densityME1}) and (\ref{eq:distME}) are valid for each $\mathrm{ME}$ density $f_j$, it is straightforward to see that (\ref{eq:residualcom1MMem1}) and (\ref{eq:residualcom2MMem1}) follow from the above definitions. From (\ref{eq:residualcom2MMem1}), it is clear that $\sum_{\bm{i}\in\mathscr{S}} p^\mathrm{RL}_{\bm{z},\bm{i}}=1$. Since each $f^\mathrm{RL}_{z,j}$ in (\ref{eq:residualcom1MMem1}) is an $\mathrm{ME}$ density, the result follows.
\end{proof}\vspace{-.3in}

\subsection{Size-biased Esscher transform}\label{subsec:SBET}\vspace{-.2in}

Fix $n_1,\dots,n_M\in\mathds{N}_0$ and $\lambda_1,\dots,\lambda_M\ge 0$. For later use, it will be convenient to collect such information in the vectors $\bm{n}=(n_1,\dots,n_M)$ and $\bm{\lambda}=(\lambda_1,\dots,\lambda_M)$. The density $f^\mathrm{ET}_{\bm{n},\bm{\lambda}}$ of a \emph{size-biased Esscher transform} of an $M$-dimensional multivariate density function $f$ with support on $[0,\infty)^M$ is defined by
\begin{equation}\label{eq:densityGenEsscher}
f^\mathrm{ET}_{\bm{n},\bm{\lambda}}(x_1,\dots,x_M)= \frac{1}{C^\mathrm{ET}_{\bm{n},\bm{\lambda}}}x_1^{n_1}\cdots x_M^{n_M} e^{-\lambda_1 x_1 \cdots - \lambda_M x_M}f(x_1,\dots,x_M)
\end{equation}
for $x_1,\dots,x_M\ge 0$, where
\begin{equation}\label{eq:fullCEsscher}
C^\mathrm{ET}_{\bm{n},\bm{\lambda}}=\int_0^\infty\cdots\int_0^\infty y_1^{n_1}\cdots y_M^{n_M} e^{-\lambda_1 y_1 \cdots - \lambda_M y_M}f(y_1,\dots,y_M)\dd y_1\cdots\dd y_M
\end{equation}
provided $C^\mathrm{ET}_{\bm{n},\bm{\lambda}}<\infty$ (which is true if $f\in\mathrm{MMEam}$). See e.g. \cite[Section 2.2]{willmot2015some}. Note that the case $\lambda_1=\dots =\lambda_M=0$ represents the multivariate cross-moment or size-biased distribution (e.g. \cite[Equation (2)]{navarro2006MultiSizeBiased}) while the case $n_1=\dots=n_M=0$ corresponds to the usual multivariate Esscher transform. In what follows we would like to prove that if $f\in\mathrm{MMEam}$, then $f^\mathrm{ET}_{\bm{n},\bm{\lambda}}\in\mathrm{MMEam}$. To do so, we first need the following result.
\begin{Lemma}\label{lem:momentdist1} 
Given a square matrix $\bm{T}$ and a real number $\lambda\in\mathds{R}$, we have that, for all $n\in \mathds{N}_0$ and $x\ge 0$,
\begin{equation}\label{eq:esscher1}
\frac{x^n}{n!}e^{-\lambda x} e^{\bm{T} x}= 
\begin{pmatrix}\bm{I}&\bm{0}&\cdots&\bm{0}\end{pmatrix}
\exp\big( \bm{T}^{[n,\lambda]} x\big)\begin{pmatrix}\bm{0}\\\vdots\\\bm{0}\\\bm{I}\end{pmatrix},
\end{equation}
where
\[\bm{T}^{[n,\lambda]}=\underbrace{\left(\begin{matrix}\bm{T}-\lambda\bm{I}& \bm{I} & &  &\\ &\bm{T}-\lambda\bm{I}& \bm{I} &  &\\ && \bm{T}-\lambda\bm{I} &\\ &&& \ddots &\\&&&&\bm{T}-\lambda\bm{I}\end{matrix}\right)}_{n+1 \;\mathrm{blocks}}.\]
\end{Lemma}\vspace{-.2in}
\begin{proof}
Employing the results of \cite{van1978computing} as in the proof of Proposition \ref{prop:convolutionME1}, we get that the right-hand side of (\ref{eq:esscher1}) is equal to 
\begin{align*}
&\int_0^x\int_0^{s_1} \cdots \int_0^{s_{n-1}} e^{(\bm{T}-\lambda\bm{I})(x-s_1)}\bm{I} e^{(\bm{T}-\lambda\bm{I})(s_1-s_2)}\bm{I} \cdots e^{(\bm{T}-\lambda\bm{I})s_{n-1}}\dd s_{n}\cdots\dd s_2\dd s_1\\
&\quad = e^{(\bm{T}-\lambda\bm{I})x}\int_0^x\int_0^{s_1} \cdots \int_0^{s_{n-1}} \dd s_{n}\cdots\dd s_2\dd s_1 =e^{-\lambda x}e^{\bm{T}x} \frac{x^n}{n!},
\end{align*}
where in the last equality we have used that $e^{-\lambda x}e^{\bm{T}x} = e^{-\lambda\bm{I} x}e^{\bm{T}x} = e^{(\bm{T}-\lambda\bm{I})x}$.
\end{proof}
\begin{theorem} \label{th:SBET} (\textbf{Size-biased Esscher transform of $\mathrm{MMEam}$}) Let $f\in\mathrm{MMEam}$ be of the form (\ref{eq:densityMMEm1}). Then, the size-biased Esscher transformed density $f^\mathrm{ET}_{\bm{n},\bm{\lambda}}$ defined in (\ref{eq:densityGenEsscher}) belongs to $\mathrm{MMEam}$ and its explicit formula is given by
\begin{equation}\label{eq:equilibriumaux1}
f^\mathrm{ET}_{\bm{n},\bm{\lambda}}(x_1,\dots,x_M)=\sum_{\bm{i}\in\mathscr{S}}p_{\bm{n},\bm{\lambda},\bm{i}}^\mathrm{ET} \,f^\mathrm{ET}_{n_1,\lambda_1,i_1}(x_1)\cdots  f^\mathrm{ET}_{n_M,\lambda_M,i_M}(x_M),
\end{equation}
where
\begin{equation}\label{eq:equilibriumaux2}
f^\mathrm{ET}_{n,\lambda,j}(x)=\left(\frac{n!}{C^\mathrm{ET}_{n,\lambda,j}}\bm{\alpha}_{j}^{[n]}\right) e^{\bm{T}_{j}^{[n, \lambda]}x} \bm{t}_{j}^{[n]},\quad C^\mathrm{ET}_{n,\lambda,j}=n!\bm{\alpha}_{j}(\lambda\bm{I}-\bm{T}_{j})^{-(n+1)}\bm{t}_{j}
\end{equation}
for $n\in\mathds{N}_0$, $\lambda\ge0$ and $1\le j \le L$, with
\begin{equation}\label{eq:alphantn1}\bm{\alpha}_j^{[n]}=\underbrace{\left(\bm{\alpha}_{j},\bm{0},\dots,\bm{0}\right)}_{n+1 \;\mathrm{blocks}}, \quad \bm{t}_j^{[n]}=\left.\begin{pmatrix}\bm{0}\\\vdots\\\bm{0}\\ \bm{t}_{j}\end{pmatrix}\right\}{\text{\scriptsize $n+1$ blocks}}.\end{equation}
In addition, the weights are defined by
\begin{equation}\label{eq:pEsscher}
p_{\bm{n},\bm{\lambda},\bm{i}}^\mathrm{ET}= p_{\bm{i}}\frac{C^\mathrm{ET}_{n_1,\lambda_1,i_1}C^\mathrm{ET}_{n_2,\lambda_2,i_2}\cdots C^\mathrm{ET}_{n_M,\lambda_M,i_M}}{C^\mathrm{ET}_{\bm{n},\bm{\lambda}}},\quad C^\mathrm{ET}_{\bm{n},\bm{\lambda}}=\sum_{\bm{i}\in\mathscr{S}} p_{\bm{i}} \,C^\mathrm{ET}_{n_1,\lambda_1,i_1}C^\mathrm{ET}_{n_2,\lambda_2,i_2}\cdots C^\mathrm{ET}_{n_M,\lambda_M,i_M}.
\end{equation}
\end{theorem}\vspace{-.2in}
\begin{proof} Substitution of (\ref{eq:densityMMEm1}) into (\ref{eq:densityGenEsscher}) yields
\[
f^\mathrm{ET}_{\bm{n},\bm{\lambda}}(x_1,\dots,x_M) = \frac{1}{C^\mathrm{ET}_{\bm{n},\bm{\lambda}}}\sum_{\bm{i}\in\mathscr{S}}p_{\bm{i}}\,\left[x_1^{n_1} e^{-\lambda_1 x_1} f_{i_1}(x_1)\right]\cdots \left[x_M^{n_M} e^{-\lambda_M x_M} f_{i_M}(x_M)\right]
\]
which equals (\ref{eq:equilibriumaux1}) under the definitions
\begin{equation}\label{eq:EsscherdensityME}
f^\mathrm{ET}_{n,\lambda,j}(x)=\frac{1}{C^\mathrm{ET}_{n,\lambda,j}}x^{n}e^{-\lambda x} f_{j}(x),\quad C^\mathrm{ET}_{n,\lambda,j}=\int_0^\infty y^{n}e^{-\lambda y} f_{j}(y)\dd y,
\end{equation}
and $p_{\bm{n},\bm{\lambda},\bm{i}}^\mathrm{ET}$ in (\ref{eq:pEsscher}). Note that $C^\mathrm{ET}_{n,\lambda,j}$ is a normalizing constant such that $f^\mathrm{ET}_{n,\lambda,j}$ is a univariate size-biased Esscher transformed density of $f_j$. With $f_j\in \mathrm{ME}$, the density $f^\mathrm{ET}_{n,\lambda,j}$ can be expressed as
\[
f^\mathrm{ET}_{n,\lambda,j}(x)=\frac{1}{C^\mathrm{ET}_{n,\lambda,j}}\bm{\alpha}_{j}\left(x^{n}e^{-\lambda x} e^{\bm{T}_{j} x}\right)\bm{t}_{j},
\]
and therefore application of Lemma \ref{lem:momentdist1} results in the representation of $f^\mathrm{ET}_{n,\lambda,j}$ in (\ref{eq:equilibriumaux2}) with $\bm{\alpha}_j^{[n]}$ and $\bm{t}_j^{[n]}$ given by (\ref{eq:alphantn1}). The constant $C^\mathrm{ET}_{n,\lambda,j}$ in (\ref{eq:EsscherdensityME}) corresponds to the $n$-th moment of a (possibly defective) $\mathrm{ME}$ distribution with parameters $(\bm{\alpha}_{j},\bm{T}_{j}-\lambda\bm{I},\bm{t}_{j})$, and thus, using \cite[Theorem 4.1.18]{bladt2017matrix} we get its explicit form in (\ref{eq:equilibriumaux2}). Then, under (\ref{eq:densityMMEm1}) the constant $C^\mathrm{ET}_{\bm{n},\bm{\lambda}}$ appearing in (\ref{eq:fullCEsscher}) is easily obtained as in (\ref{eq:pEsscher}). Finally, (\ref{eq:equilibriumaux2}) implies that $f^\mathrm{ET}_{n,\lambda,j}\in\mathrm{ME}$ for each $j$ while (\ref{eq:pEsscher}) means that $\sum_{\bm{i}\in\mathscr{S}}p_{\bm{n},\bm{\lambda},\bm{i}}^\mathrm{ET}=1$. Therefore, $f^\mathrm{ET}_{\bm{n},\bm{\lambda}}\in\mathrm{MMEam}$ according to the representation (\ref{eq:equilibriumaux1}). 
\end{proof}
\begin{Remark}\label{rem:Esschertransform}
In the above theorem, the Esscher transform arguments $\lambda_1,\dots,\lambda_M$ are assumed non-negative for the ease of exposition. If we define $\kappa_j<0$ to be the dominant eigenvalue of $\bm{T}_j$ (see Section \ref{sec:ME}), then our results are still valid as long as $\lambda_i>\kappa_j$ for $1\le i\le M$ and $1\le j \le L$, as this guarantees that the integral representation of $C^\mathrm{ET}_{n,\lambda,j}$ in (\ref{eq:EsscherdensityME}) is finite and the density in (\ref{eq:EsscherdensityME}) is well-defined for $\lambda=\lambda_1,\dots,\lambda_M$ and $1\le j \le L$.
\end{Remark}
\vspace{-.3in}

\subsection{Higher order equilibrium distribution}\label{subsec:EqmDist}\vspace{-.2in}
Given $r\in\mathds{N}_+$ and an $M$-variate density function $f$ with support on $[0,\infty)^M$, let us recursively define the density
\begin{equation}\label{eq:equilibriumdistGen}
f^\mathrm{ED}_{r}(x_1,\dots,x_M)= \frac{\overline{F}^\mathrm{ED}_{r-1}(x_1,\dots,x_M)}{\int_0^\infty \cdots\int_0^\infty
\overline{F}^\mathrm{ED}_{r-1}(y_1,\dots,y_M)\dd y_1\cdots\dd y_M}
\end{equation}
for $x_1,\dots,x_M\ge 0$, where $\overline{F}^\mathrm{ED}_{r}(x_1,\dots,x_M)=\int_{x_M}^\infty\cdots\int_{x_1}^\infty f^\mathrm{ED}_{r}(u_1,\dots,u_M)\dd u_1\cdots \dd u_M$ is the corresponding survival function and $f^\mathrm{ED}_0=f$ is the starting point  (and hence $\overline{F}^\mathrm{ED}_0=\overline{F}$). The function $f^\mathrm{ED}_r$ is called the \emph{$r$-th order equilibrium density function} of $f$ (see e.g. \cite[Equation (23)]{willmot2015some}).
\begin{theorem} (\textbf{Higher order equilibrium distribution of $\mathrm{MMEam}$})
Suppose that $f\in\mathrm{MMEam}$ is of the form (\ref{eq:densityMMEm1}). Then, for all $r\in\mathds{N}_+$, the $r$-th order equilibrium density function $f^\mathrm{ED}_r$ defined in (\ref{eq:equilibriumdistGen}) belongs to $\mathrm{MMEam}$, and is of the form
\begin{equation}\label{eq:equilibriumdist1}
f^\mathrm{ED}_r(x_1,\dots,x_M) =\sum_{\bm{i}\in\mathscr{S}} p^\mathrm{ED}_{r,\bm{i}}\,f^\mathrm{ED}_{r,i_1}(x_1)\cdots f^\mathrm{ED}_{r,i_M}(x_M),
\end{equation}
where
\begin{equation}\label{eq:Eqmcom1}
f^\mathrm{ED}_{r,j}(x) = \bm{\alpha}^\mathrm{ED}_{r,j}e^{\bm{T}_j x}\bm{l}_j,\quad \bm{\alpha}^\mathrm{ED}_{r,j}= \frac{1}{\bm{\alpha}^\mathrm{ED}_{r-1,j}(-\bm{T}_j)^{-1}\bm{l}_j}\bm{\alpha}^\mathrm{ED}_{r-1,j}
\end{equation}
for $1\le j \le L$ and
\begin{equation}\label{eq:Eqmcom2}
p^\mathrm{ED}_{r,\bm{i}}=\frac{p^\mathrm{ED}_{r-1,\bm{i}} \left(\bm{\alpha}^\mathrm{ED}_{r-1,i_1}(-\bm{T}_{i_1})^{-1}\bm{l}_{i_1}\right)\cdots\left(\bm{\alpha}^\mathrm{ED}_{r-1,i_M}(-\bm{T}_{i_M})^{-1}\bm{l}_{i_M}\right)}{\sum_{\bm{h}\in\mathscr{S}} p^\mathrm{ED}_{r-1,\bm{h}}\left(\bm{\alpha}^\mathrm{ED}_{r-1,h_1}(-\bm{T}_{h_1})^{-1}\bm{l}_{h_1}\right)\cdots\left(\bm{\alpha}^\mathrm{ED}_{r-1,h_M}(-\bm{T}_{h_M})^{-1}\bm{l}_{h_M}\right)}
\end{equation}
are obtained recursively with the starting point $\bm{\alpha}^\mathrm{ED}_{0,j}=\bm{\alpha}_{j}$ and $p^\mathrm{ED}_{0,\bm{i}}=p_{\bm{i}}$.
\end{theorem}\vspace{-.2in}
\begin{proof}
By substituting (\ref{eq:densityMMEm1}), we have the first order equilibrium density
\begin{align*}
f^\mathrm{ED}_1(x_1,\dots,x_M) & = \frac{\sum_{\bm{i}\in\mathscr{S}} p_{\bm{i}}\overline{F}_{i_1}(x_1)\cdots\overline{F}_{i_M}(x_M)}{\sum_{\bm{h}\in\mathscr{S}} p_{\bm{h}}\left(\int_0^\infty \overline{F}_{h_1}(y)\dd y\right)\cdots\left(\int_0^\infty \overline{F}_{h_M}(y)\dd y\right)}\\
& = \sum_{\bm{i}\in\mathscr{S}} p^\mathrm{ED}_{1,\bm{i}}\,f^\mathrm{ED}_{1,i_1}(x_1)\cdots f^\mathrm{ED}_{1,i_M}(x_M),
\end{align*}
where
\begin{align*}
f^\mathrm{ED}_{1,j}(x) = \frac{\overline{F}_j(x)}{\int_0^\infty \overline{F}_j(y)\dd y}=\frac{\bm{\alpha}_je^{\bm{T}_j x}\bm{l}_j}{\bm{\alpha}_j(-\bm{T}_j)^{-1}\bm{l}_j}= \bm{\alpha}^\mathrm{ED}_{1,j}e^{\bm{T}_j x}\bm{l}_j,\quad \bm{\alpha}^\mathrm{ED}_{1,j}= \frac{1}{\bm{\alpha}_j(-\bm{T}_j)^{-1}\bm{l}_j}\bm{\alpha}_{j}
\end{align*}
for $1\le j \le L$, and
\begin{align*}
p^\mathrm{ED}_{1,\bm{i}} & = \frac{p_{\bm{i}} \left(\int_0^\infty \overline{F}_{i_1}(y)\dd y\right)\cdots\left(\int_0^\infty \overline{F}_{i_M}(y)\dd y\right)}{\sum_{\bm{h}\in\mathscr{S}} p_{\bm{h}}\left(\int_0^\infty \overline{F}_{h_1}(y)\dd y\right)\cdots\left(\int_0^\infty \overline{F}_{h_M}(y)\dd y\right)}\\
& =\frac{p_{\bm{i}} \left(\bm{\alpha}_{i_1}(-\bm{T}_{i_1})^{-1}\bm{l}_{i_1}\right)\cdots\left(\bm{\alpha}_{i_M}(-\bm{T}_{i_M})^{-1}\bm{l}_{i_M}\right)}{\sum_{\bm{h}\in\mathscr{S}} p_{\bm{h}}\left(\bm{\alpha}_{h_1}(-\bm{T}_{h_1})^{-1}\bm{l}_{h_1}\right)\cdots\left(\bm{\alpha}_{h_M}(-\bm{T}_{h_M})^{-1}\bm{l}_{h_M}\right)}.
\end{align*}
Note that we have used that $\int_0^\infty \overline{F}_{j}(y)\dd y=\bm{\alpha}_j\left(\int_0^\infty e^{\bm{T}_j y}\dd y\right)\bm{l}_j = \bm{\alpha}_j(-\bm{T}_j)^{-1}\bm{l}_j$.
Recursively, it is straightforward to see that (\ref{eq:equilibriumdist1})-(\ref{eq:Eqmcom2}) are valid when $r\ge 2$. Clearly, the results also hold true for $r=1$ by defining $\bm{\alpha}^\mathrm{ED}_{0,j}=\bm{\alpha}_{j}$ and $p^\mathrm{ED}_{0,\bm{i}}=p_{\bm{i}}$. Since $f^\mathrm{ED}_{r,j}\in \mathrm{ME}$ for each $j\in\{1,\dots,L\}$ and $r\in\mathds{N}_+$, we have that $f^\mathrm{ED}_{r}\in \mathrm{MMEam}$ for $r\in\mathds{N}_+$. We remark that $f^\mathrm{ED}_{r,j}$ is the $r$-th order univariate equilibrium density of $f_j$.
\end{proof}\vspace{-.3in}

\subsection{Order statistics}\label{sec:OS}\vspace{-.2in}
For a random vector $(X_1, \dots, X_M)$ that follows the $M$-variate density function $f$ with support on $[0,\infty)^M$, we denote by $X_{j: M}^\mathrm{OS}$ the $j$-th order statistic and let $f^\mathrm{OS}_{j:M}$ be its associated density for $1\le j\le M$. It is shown below that each $f^\mathrm{OS}_{j:M}$ belongs to $\mathrm{MEam}$.
\begin{theorem}\label{th:OS} (\textbf{Order statistics of $\mathrm{MMEam}$}) Suppose that $f\in\mathrm{MMEam}$ is of the form (\ref{eq:densityMMEm1}). Then, $f^\mathrm{OS}_{j:M}\in\mathrm{MEam}$ for each $j\in\{1,\dots,M\}$.
\end{theorem}\vspace{-.2in}
\begin{proof}
Let $\mathscr{P}_M$ denote the set of permutations of $\{1,\dots, M\}$. Following the law of total probability, we have
\begin{align}
&f_{j:M}^\mathrm{OS}(x) = \mathds{P}(X_{j: M}^\mathrm{OS}\in [x,x+\dd x])/\dd x \nonumber\\
&\quad = \frac{1}{(j-1)!(M-j)!}\sum_{\bm{r}\in\mathscr{P}_M} \mathds{P}(X_{r_1} < x,\dots, X_{r_{j-1}} < x, X_{r_j}\in [x,x+\dd x], X_{r_{j+1}} > x,\dots, X_{r_{M}} > x)/\dd x \nonumber\\
&\quad = \frac{1}{(j-1)!(M-j)!}\sum_{\bm{r}\in\mathscr{P}_M} \sum_{\bm{i}\in\mathscr{S}}p_{\bm{i}}F_{i_{r_1}}(x)\cdots F_{i_{r_{j-1}}} (x) f_{i_{r_j}}(x)\overline{F}_{i_{r_{j+1}}}(x)\cdots\overline{F}_{i_{r_{M}}}(x)\nonumber\\
&\quad = \sum_{\bm{i}\in\mathscr{S}}p_{\bm{i}} f_{\bm{i},j:M}(x)\label{eq:orderm1}
\end{align}
for $x\ge 0$, where $f_{\bm{i},j:M}$ is the $j$-th order statistic of $f_{i_1}, f_{i_2},\dots, f_{i_M}$. Since the latter are elements of $\mathrm{ME}$, Proposition \ref{prop:order1} implies that $f_{\bm{i},j:M}\in \mathrm{MEam}$, so that (\ref{eq:orderm1}) is also in $\mathrm{MEam}$.
\end{proof}\vspace{-.3in}
\subsection{Correlation measures}\vspace{-.2in}

If a random vector $(X_1, \dots, X_M)$ has the $M$-variate density function $f\in\mathrm{MMEam}$ of the form (\ref{eq:densityMMEm1}), then it is straightforward to apply the formulas for (unconditional) cross moments in Theorem \ref{th:crossmoments1} to compute the Pearson's correlation coefficient between any two random variables $X_{j_1}$ and $X_{j_2}$, which is defined by $(\mathds{E}[X_{j_1} X_{j_2}]-\mathds{E}[X_{j_1}]\mathds{E}[X_{j_2}])/\sqrt{\mathrm{Var}(X_{j_1})\mathrm{Var}(X_{j_2})}$. 
Closed-form formulas for other common rank correlation measures which describe the dependency between $X_{j_1}$ and $X_{j_2}$, namely Kendall's tau and Spearman's rho, can also be obtained, and these are the subject matters of this subsection. For later use, it is understood that the entries of $\bm{i}',\bm{i}''\in\mathscr{S}$ are given by $\bm{i}'=(i_1',\dots, i_M')$ and $\bm{i}''=(i_1'',\dots, i_M'')$ respectively.\vspace{-.2in}
\subsubsection{Kendall's tau}\vspace{-.2in}
Let $(Y,Z)$ be a random vector which is assumed to follow the bivariate density $h$ and cumulative distribution function $H$. Define the coefficient 
\[\tau(Y,Z)=\mathds{P}((Y_1-Y_2)(Z_1-Z_2)>0) - \mathds{P}((Y_1-Y_2)(Z_1-Z_2)<0),\]
where $(Y_1,Z_1)$ and $(Y_2,Z_2)$ are independent copies of $(Y,Z)$. Then $\tau(Y,Z)$, called \emph{Kendall's tau}, measures the theoretical \emph{concordance} between the entries of $(Y_1,Z_1)$ and $(Y_2, Z_2)$. It was proved in \cite[Theorem 5.1.1]{nelsen2007introduction} that 
\begin{equation}\label{eq:Kendall1}\tau(Y,Z)=4\int_0^\infty\int_0^\infty H(y,z) h(y,z)\dd y \dd z-1.\end{equation}
In the following we provide a formula for $\tau(X_{j_1},X_{j_2})$ whenever the density function of $(X_1,\dots,X_M)$ belongs to $\mathrm{MMEam}$.
\begin{theorem}\label{th:Kendall1} (\textbf{Kendall's tau of $\mathrm{MMEam}$}) Suppose that $(X_1,\dots,X_M)$ follow the density $f\in\mathrm{MMEam}$ which is of the form (\ref{eq:densityMMEm1}). Then, for $1\le j_1,j_2\le M$, the Kendall's tau of the pair $(X_{j_1},X_{j_2})$ is given by
\[\tau(X_{j_1},X_{j_2})=4 \sum_{\bm{i},\bm{i}'\in\mathscr{S}} p_{\bm{i}}\,p_{\bm{i}'} \,c_{\bm{i},\bm{i}'}(j_1) \,c_{\bm{i},\bm{i}'}(j_2) -1,\]
where
\begin{equation}\label{eq:KendallcDef}
c_{\bm{i},\bm{i}'}(r)= 1-\left(\bm{\alpha}_{i_r}\otimes \bm{\alpha}_{i_r'}\right) \left(-\bm{T}_{i_r} \oplus \bm{T}_{i_r'}\right)^{-1} \left(\bm{l}_{i_r}\otimes \bm{t}_{i_r'}\right).
\end{equation}
\end{theorem}\vspace{-.2in}
\begin{proof}
With $(X_{j_1},X_{j_2})$ in place of $(Y,Z)$, the assumption (\ref{eq:densityMMEm1}) implies that
\[
h(y,z)=\sum_{\bm{i}'\in\mathscr{S}} p_{\bm{i}'} f_{i_{j_1}'}(y) f_{i_{j_2}'}(z), \quad H(y,z)=\sum_{\bm{i}\in\mathscr{S}} p_{\bm{i}} F_{i_{j_1}}(y) F_{i_{j_2}}(z).
\]
Hence, using (\ref{eq:Kendall1}), we have
\[
\tau(X_{j_1},X_{j_2}) = 4 \sum_{\bm{i},\bm{i}'\in\mathscr{S}} p_{\bm{i}}\,p_{\bm{i}'} \left(\int_0^\infty F_{i_{j_1}}(y) f_{i_{j_1}'}(y)  \dd y\right) \left(\int_0^\infty F_{i_{j_2}}(z)f_{i_{j_2}'}(z)  \dd z\right)-1.
\]
Since
\begin{align}
\int_0^\infty F_{i_r}(x) f_{i_r'}(x)  \dd x & = 1- \int_0^\infty \overline{F}_{i_r}(x) f_{i_r'}(x)\dd x\nonumber\\
& = 1-\int_0^\infty \left(\bm{\alpha}_{i_r}e^{\bm{T}_{i_r} x} \bm{l}_{i_r}\right)\left(\bm{\alpha}_{i_r'}e^{\bm{T}_{i_r'} x} \bm{t}_{i_r'}\right)\dd x \nonumber\\
& = 1-\int_0^\infty \left(\bm{\alpha}_{i_r}\otimes \bm{\alpha}_{i_r'}\right) e^{\left(\bm{T}_{i_r} \oplus \bm{T}_{i_r'}\right)x} \left(\bm{l}_{i_r}\otimes \bm{t}_{i_r'}\right)\dd x \nonumber\\
& = 1-\left(\bm{\alpha}_{i_r}\otimes \bm{\alpha}_{i_r'}\right) \left(-\bm{T}_{i_r} \oplus \bm{T}_{i_r'}\right)^{-1} \left(\bm{l}_{i_r}\otimes \bm{t}_{i_r'}\right),\label{eq:KendallProofIdentity}
\end{align}
the result follows.
\end{proof}\vspace{-.3in}
\subsubsection{Spearman's rho}\vspace{-.2in}
Suppose now that for a vector $(Y,Z)$ with joint cumulative distribution function $H$, we want to measure its concordance with respect to a version of itself with identical but independent marginals. With $(Y_1,Z_1)$, $(Y_2,Z_2)$ and $(Y_3,Z_3)$ independent copies of $(Y,Z)$, the \emph{Spearman's rho} is defined as
\[\rho(Y,Z)= 3\big(\mathds{P}((Y_1-Y_2)(Z_1-Z_3)>0) - \mathds{P}((Y_1-Y_2)(Z_1-Z_3)<0)\big).\]
As in the case of Kendall's tau, $\rho(Y,Z)$ admits an integral expression \cite[Theorem 5.1.6]{nelsen2007introduction}
\begin{equation}\label{eq:Spearman1}\rho(Y,Z)=12\int_0^\infty\int_0^\infty H(y,z)h_{Y}(y) h_{Z}(z)\dd y \dd z -3,\end{equation}
where $h_{Y}$ and $h_{Z}$ are the marginal densities of $Y$ and $Z$ respectively. Next, we provide a closed-form expression for $\rho(X_{j_1},X_{j_2})$ when $(X_1,\dots,X_M)$ is distributed as $\mathrm{MMEam}$.
\begin{theorem}{ (\textbf{Spearman's rho of $\mathrm{MMEam}$}) Suppose that $(X_1,\dots,X_M)$ follow the density $f\in\mathrm{MMEam}$ which is of the form (\ref{eq:densityMMEm1}). Then, for $1\le j_1,j_2\le M$}, 
the Spearman's rho of the pair $(X_{j_1},X_{j_2})$ is given by
\[\rho(X_{j_1},X_{j_2})= 12 \sum_{\bm{i}, \bm{i}', \bm{i}''\in\mathscr{S}}  p_{\bm{i}}\,p_{\bm{i}'}\,p_{\bm{i}''}\,c_{\bm{i},\bm{i}'}(j_1) \,c_{\bm{i},\bm{i}''}(j_2) - 3,\]
where $c_{\bm{i},\bm{i}'}(j_1)$ and $c_{\bm{i},\bm{i}''}(j_2)$ are defined via (\ref{eq:KendallcDef}).
\end{theorem}\vspace{-.2in}
\begin{proof}
With $(X_{j_1},X_{j_2})$ in place of $(Y,Z)$, the integral in the definition (\ref{eq:Spearman1}) can be evaluated as
\begin{align*}
&\int_0^\infty\int_0^\infty H(y,z)h_{Y}(y) h_{Z}(z)\dd y\dd z\\
&\quad  = \int_0^\infty\int_0^\infty \left(\sum_{\bm{i}\in\mathscr{S}} p_{\bm{i}} F_{i_{j_1}}(y) F_{i_{j_2}}(z)\right) \left(\sum_{\bm{i}'\in\mathscr{S}} p_{\bm{i}'} f_{i_{j_1}'}(y)\right) \left(\sum_{\bm{i}''\in\mathscr{S}} p_{\bm{i}''} f_{i_{j_2}''}(z)\right) \dd y \dd z\\
&\quad = \sum_{\bm{i}, \bm{i}', \bm{i}''\in\mathscr{S}} p_{\bm{i}}\,p_{\bm{i}'}\,p_{\bm{i}''}\left(\int_0^\infty F_{i_{j_1}}(y) f_{i_{j_1}'}(y)  \dd y\right) \left(\int_0^\infty F_{i_{j_2}}(z) f_{i_{j_2}''}(z)  \dd z\right) \\
&\quad = \sum_{\bm{i}, \bm{i}', \bm{i}''\in\mathscr{S}}  p_{\bm{i}}\,p_{\bm{i}'}\,p_{\bm{i}''}\,c_{\bm{i},\bm{i}'}(j_1) \,c_{\bm{i},\bm{i}''}(j_2),
\end{align*}
where the last equality follows from (\ref{eq:KendallProofIdentity}), and the proof is complete.
\end{proof}\vspace{-.3in}
\section{Applications to risk theory}\label{sec:Applications}\vspace{-.2in}

In this section, we shall apply the results in previous sections to solve various problems in risk theory. Unless specified otherwise, it is assumed that an insurer has a portfolio of $M$ dependent risks (or losses) represented by the non-negative random vector $\bm{X}=(X_1,\dots, X_M)$ which follows the density $f\in\mathrm{MMEam}$ in (\ref{eq:densityMMEm1}).\vspace{-.2in}

\subsection{Multivariate excess loss, tail conditional expectation and tail covariance}\label{subsec:ELandMCTE}\vspace{-.2in}

For a given $j$-th risk $X_j$, the (univariate) residual lifetime $X_j-z_j|X_j>z_j$ corresponds to its excess loss random variable under a \emph{deductible} of $z_j\ge0$. With the notation $y_+=\max(y,0)$, this relates to an \emph{excess-of-loss (re)insurance} contract where the (re)insurer will be responsible for the amount $(X_j-z_j)_+$ whose expectation equals $\mathds{E}[(X_j-z_j)_+] = \mathds{E}[X_j-z_j|X_j>z_j]\overline{F}_{X_j}(z_j)$. Here $\overline{F}_{X_j}$ is the survival function of $X_j$ and $\mathds{E}[X_j-z_j|X_j>z_j]$ is known as the \emph{mean excess loss function} (see also Remark \ref{rem:SLtransform}). In the present multivariate context, the residual lifetime of $\bm{X}-\bm{z}$ conditional on $\bm{X}>\bm{z}$ (where $\bm{z}=(z_1,\dots,z_M)$) considered in Section \ref{sec:RL} may also be interpreted as the multivariate excess loss. In the following we provide closed-form formulas for the higher order cross moments.
\begin{theorem}\label{th:crossmomentsRL} (\textbf{Cross moments of excess losses of $\mathrm{MMEam}$}) Suppose that $(X_1, \dots, X_M)$ follows the density $f\in\mathrm{MMEam}$ of the form (\ref{eq:densityMMEm1}), and let $z_{1}, \dots , z_{M}\ge 0$ and $r_1,\dots,r_M\in\mathds{N}_0$. Then,
\begin{equation}\label{eq:MomentEL}
\mathds{E}\left[\left.\prod_{j = 1}^M (X_j-z_j)^{r_j}\;\right| \;X_{1}>z_{1},\dots, X_M>z_M\right] = \sum_{\bm{i}\in\mathscr{S}} p^{\mathrm{RL}}_{\bm{z},\bm{i}} \prod_{j = 1}^M r_j!\, \bm{\alpha}^\mathrm{RL}_{z_j,i_j}(-\bm{T}_{i_j})^{-(r_j+1)}\bm{t}_{i_j},
\end{equation}
where $p^{\mathrm{RL}}_{\bm{z},\bm{i}}$ and $\bm{\alpha}^\mathrm{RL}_{z_j,i_j}$ are defined as in Theorem \ref{th:residual1}.
\end{theorem}\vspace{-.2in}
\begin{proof} By Theorem \ref{th:residual1}, the density of the vector $\bm{X}-\bm{z}\;|\;\bm{X}>\bm{z}$ belongs to $\mathrm{MMEam}$ and is of the form (\ref{eq:residualMMem1}). The result follows by applying Theorem \ref{th:crossmoments1} with $k=M$.
\end{proof}\vspace{-.2in}

Excess loss is closely related to the notion of \emph{tail conditional expectation} ($\mathrm{TCE}$) which is an important risk measure (e.g. \cite{ArtznerMF1999, goovaerts2010RM}). We first recall that, in the univariate setting, the \emph{Value-at-Risk} ($\mathrm{V@R}$)  at level $\theta\in[0,1)$ for a loss random variable $Y$ with cumulative distribution function $H_Y$ is defined by $\mathrm{V@R}_\theta(Y)=\inf\{y:H_Y(y)\ge \theta\}$. Then, the $\mathrm{TCE}$ at level $\theta$ is defined by the expectation $\mathrm{TCE}_\theta(Y)=\mathds{E}[Y|Y>\mathrm{V@R}_\theta(Y)]$. Clearly, $\mathrm{TCE}_\theta(Y)=\mathrm{V@R}_\theta(Y)+\mathds{E}[Y-\mathrm{V@R}_\theta(Y)| Y>\mathrm{V@R}_\theta(Y)]$, where the expectation is the mean excess loss over $\mathrm{V@R}_\theta(Y)$. When $Y$ is a continuous random variable, $\mathrm{TCE}_\theta(Y)$ coincides with the Tail-Value-at-Risk ($\mathrm{TV@R}$) defined via $\mathrm{TV@R}_\theta(Y)=\int_\theta^1\mathrm{V@R}_u(Y)\dd u/(1-\theta)$. $\mathrm{TV@R}$ is also known as \emph{expected shortfall} ($\mathrm{ES}$).

In \cite{landsman2016multivariate, landsman2018multivariate}, a multivariate risk measure based on the first moment of $\bm{X}$ conditional on $\bm{X}>\mathrm{V@R}_{\bm{\theta}}(\bm{X})$ was proposed, where $\mathrm{V@R}_{\bm{\theta}}(\bm{X}) = (\mathrm{V@R}_{\theta_1}(X_1),\dots,\mathrm{V@R}_{\theta_M}(X_M))$. Here the quantiles are collected in the vector $\bm{\theta}=(\theta_1,\dots,\theta_M)$ with $\theta_j\in [0,1)$ for $1\le j\le M$. Such a risk measure is called \emph{multivariate tail conditional expectation} ($\mathrm{MTCE}$) and is defined by the $M$-dimensional vector
\begin{equation}\label{eq:MTCEdef}
\mathrm{MTCE}_{\bm{\theta}}(\bm{X}) = \mathds{E}[\bm{X}|\bm{X}>\mathrm{V@R}_{\bm{\theta}}(\bm{X})],
\end{equation}
where its $j$-th component is given by $\mathds{E}[X_j|\bm{X}>\mathrm{V@R}_{\bm{\theta}}(\bm{X})]$. Note from (\ref{eq:densityMMEm1}) that each $X_j$ follows an $\mathrm{MEam}$ distribution with continuous cumulative distribution function given by
\begin{equation}\label{FXjDef}
F_{X_j}(x) = \sum_{\bm{i}\in\mathscr{S}}p_{\bm{i}} F_{i_j}(x) = 1- \sum_{\bm{i}\in\mathscr{S}}p_{\bm{i}} \bm{\alpha}_{i_j}e^{\bm{T}_{i_j}x}\bm{l}_{i_j}
\end{equation}
for $x\ge0$, and therefore $\mathrm{V@R}_{\theta_j}(X_j)$ can be computed from $F_{X_j}(\mathrm{V@R}_{\theta_j}(X_j))=\theta_j$. In addition to $\mathds{E}[X_j|\bm{X}>\mathrm{V@R}_{\bm{\theta}}(\bm{X})]$ which is the conditional mean of the $j$-th risk, the variability of the risk at the tail and its covariance with other risks are also of the insurer's interest. Following \cite{landsman2018multivariate}, the \emph{multivariate tail covariance} ($\mathrm{MTCov}$) is defined by the $M$-dimensional square matrix
\begin{equation}\label{eq:MTCovdef}
\mathrm{MTCov}_{\bm{\theta}}(\bm{X}) = \mathds{E}[(\bm{X}-\mathrm{MTCE}_{\bm{\theta}}(\bm{X}) )^\top (\bm{X}-\mathrm{MTCE}_{\bm{\theta}}(\bm{X}) )|\bm{X}>\mathrm{V@R}_{\bm{\theta}}(\bm{X})],
\end{equation}
where `$\top$' denotes the transpose of a matrix. For $1\le j_1,j_2\le M$, the $(j_1,j_2)$-th element of $\mathrm{MTCov}_{\bm{\theta}}(\bm{X})$ is 
\begin{align}
&\mathds{E}[(X_{j_1}-\mathds{E}[X_{j_1}|\bm{X}>\mathrm{V@R}_{\bm{\theta}}(\bm{X})])(X_{j_2}-\mathds{E}[X_{j_2}|\bm{X}>\mathrm{V@R}_{\bm{\theta}}(\bm{X})])|\bm{X}>\mathrm{V@R}_{\bm{\theta}}(\bm{X})]\nonumber\\
&\quad  = \mathds{E}[X_{j_1}X_{j_2}|\bm{X}>\mathrm{V@R}_{\bm{\theta}}(\bm{X})] - \mathds{E}[X_{j_1}|\bm{X}>\mathrm{V@R}_{\bm{\theta}}(\bm{X})] \mathds{E}[X_{j_2}|\bm{X}>\mathrm{V@R}_{\bm{\theta}}(\bm{X})].\label{eq:MTCovComponent}
\end{align}
The next theorem can be obtained by utilizing Theorem \ref{th:crossmomentsRL}.
\begin{theorem}\label{th:tailconditional1} (\textbf{$\mathrm{MTCE}$ and $\mathrm{MTCov}$ of $\mathrm{MMEam}$}) Suppose that $(X_1, \dots, X_M)$ follows the density $f\in\mathrm{MMEam}$ of the form (\ref{eq:densityMMEm1}). Then, for $1\le j\le M$, the $j$-th element of the MTCE defined in (\ref{eq:MTCEdef}) under $\theta_1,\dots,\theta_M\in [0,1)$ is given by
\begin{equation}\label{eq:MTCEresult}
\mathds{E}[X_j|\bm{X}>\mathrm{V@R}_{\bm{\theta}}(\bm{X})] = \mathrm{V@R}_{\theta_j}(X_j) + \sum_{\bm{i}\in\mathscr{S}} p^{\mathrm{RL}}_{\mathrm{V@R}_{\bm{\theta}}(\bm{X}) ,\bm{i}} \bm{\alpha}^\mathrm{RL}_{\mathrm{V@R}_{\theta_j}(X_j),i_j}(-\bm{T}_{i_j})^{-2}\bm{t}_{i_j}.
\end{equation}
Concerning the $(j_1,j_2)$-th element (\ref{eq:MTCovComponent}) of $\mathrm{MTCov}_{\bm{\theta}}(\bm{X})$ defined in (\ref{eq:MTCovdef}), the term $\mathds{E}[X_{j_1}X_{j_2}|\bm{X}>\mathrm{V@R}_{\bm{\theta}}(\bm{X})]$ depends on whether $j_1=j_2$. In particular, for $1\le j_1,j_2\le M$, one has
\begin{align}
& \mathds{E}[X_{j_1}X_{j_2}|\bm{X}>\mathrm{V@R}_{\bm{\theta}}(\bm{X})] = \mathds{E}[X_{j}^2|\bm{X}>\mathrm{V@R}_{\bm{\theta}}(\bm{X})]\nonumber\\
&\quad =-(\mathrm{V@R}_{\theta_j}(X_j))^2 + 2 \mathrm{V@R}_{\theta_j}(X_j) \mathds{E}[X_j|\bm{X}>\mathrm{V@R}_{\bm{\theta}}(\bm{X})] + 2\sum_{\bm{i}\in\mathscr{S}} p^{\mathrm{RL}}_{\mathrm{V@R}_{\bm{\theta}}(\bm{X}) ,\bm{i}} \bm{\alpha}^\mathrm{RL}_{\mathrm{V@R}_{\theta_j}(X_j),i_j}(-\bm{T}_{i_j})^{-3}\bm{t}_{i_j}\label{eq:MTCov1}
\end{align}
for $j_1=j_2=j$, and
\begin{align}
& \mathds{E}[X_{j_1}X_{j_2}|\bm{X}>\mathrm{V@R}_{\bm{\theta}}(\bm{X})]\nonumber\\
&\quad =-\mathrm{V@R}_{\theta_{j_1}}(X_{j_1})\mathrm{V@R}_{\theta_{j_2}}(X_{j_2}) +  \mathrm{V@R}_{\theta_{j_1}}(X_{j_1}) \mathds{E}[X_{j_2}|\bm{X}>\mathrm{V@R}_{\bm{\theta}}(\bm{X})] \nonumber\\
& \quad\quad\, +\mathrm{V@R}_{\theta_{j_2}}(X_{j_2}) \mathds{E}[X_{j_1}|\bm{X}>\mathrm{V@R}_{\bm{\theta}}(\bm{X})]\nonumber\\
& \quad\quad\,+ \sum_{\bm{i}\in\mathscr{S}} p^{\mathrm{RL}}_{\mathrm{V@R}_{\bm{\theta}}(\bm{X}) ,\bm{i}} \left(\bm{\alpha}^\mathrm{RL}_{\mathrm{V@R}_{\theta_{j_1}}(X_{j_1}),i_{j_1}}(-\bm{T}_{i_{j_1}})^{-2}\bm{t}_{i_{j_1}}\right) \left(\bm{\alpha}^\mathrm{RL}_{\mathrm{V@R}_{\theta_{j_2}}(X_{j_2}),i_{j_2}}(-\bm{T}_{i_{j_2}})^{-2}\bm{t}_{i_{j_2}}\right) \label{eq:MTCov2}
\end{align}
for $j_1\ne j_2$. The terms involving $\mathds{E}[X_j|\bm{X}>\mathrm{V@R}_{\bm{\theta}}(\bm{X})]$ (with $j=j_1$ or $j=j_2$) appearing in (\ref{eq:MTCovComponent}), (\ref{eq:MTCov1}) and (\ref{eq:MTCov2}) can be computed using (\ref{eq:MTCEresult}).
\end{theorem}\vspace{-.2in}
\begin{proof} 
First, we note that, for $z\ge0$ and $1\le j\le M$, the quantity $\bm{\alpha}^\mathrm{RL}_{z,j}e^{\bm{T}_j x} \bm{t}_j$ in (\ref{eq:residualcom1MMem1}) is the (univariate) residual lifetime density which must integrate to one and hence $\bm{\alpha}^\mathrm{RL}_{z,j}(-\bm{T}_j)^{-1}\bm{t}_j=1$. The result (\ref{eq:MTCEresult}) concerning $\mathrm{MTCE}$ follows from Theorem \ref{th:crossmomentsRL} by letting $r_k=\mathds{1}\{k=j\}$ and $z_k=\mathrm{V@R}_{\theta_k}(X_k)$ for $1\le k\le M$. To show (\ref{eq:MTCov1}), we note that, for $j_1=j_2=j$,
\[
\mathds{E}[X_j^2|\bm{X}>\bm{z}] =-z_j^2 + 2z_j \mathds{E}[X_j|\bm{X}>\bm{z}] + \mathds{E}[(X_j-z_j)^2|\bm{X}>\bm{z}].
\]
Letting $r_k=2\mathds{1}\{k=j\}$ and $z_k=\mathrm{V@R}_{\theta_k}(X_k)$ for $1\le k\le M$ and using Theorem \ref{th:crossmomentsRL} again, one arrives at the result (\ref{eq:MTCov1}). Similarly, when $j_1\ne j_2$, the result (\ref{eq:MTCov2}) is a direct consequence of the identity
\[
\mathds{E}[X_{j_1}X_{j_2}|\bm{X}>\bm{z}] =-z_{j_1}z_{j_2} + z_{j_1} \mathds{E}[X_{j_2}|\bm{X}>\bm{z}] + z_{j_2} \mathds{E}[X_{j_1}|\bm{X}>\bm{z}] + \mathds{E}[(X_{j_1}-z_{j_1})(X_{j_2}-z_{j_2})|\bm{X}>\bm{z}],
\]
and the proof is complete.
\end{proof}

\begin{Remark}\label{rem:SLtransform}
In a univariate context, for $r\in\mathds{N}_+$ and $z_j\ge0$, the quantity $\mathds{E}[(X_j-z_j)_+^r]$ is known as the \emph{$r$-th stop-loss transform} of the $j$-th risk. While there is a well known identity linking $\mathds{E}[(X_j-z_j)_+^r]$ and the $r$-th order equilibrium distribution of $X_j$ (e.g. \cite[Definition 1]{Hesselager1997Equi}), it is worthwhile to point out that a multivariate version also exists (see \cite[Theorem 4]{Nair2008MultiEqm}). Using our notation in Section \ref{subsec:EqmDist}, this can be stated as, for $r\in\mathds{N}_+$ and $z_{1}, \dots , z_{M}\ge 0$,
\[
\mathds{E}\left[\prod_{j = 1}^M (X_j-z_j)_+^{r}\right] =
\mathds{E}\left[\left.\prod_{j = 1}^M (X_j-z_j)^{r}\;\right| \;\bm{X}>\bm{z}\right] \overline{F}(z_1,\dots,z_M) =\mathds{E}\left[\prod_{j = 1}^M X_j^{r}\right] \overline{F}^\mathrm{ED}_{r}(z_1,\dots,z_M),
\]
where the expectation in the second expression above is linked to the left-hand side of (\ref{eq:MomentEL}) with $r_j=r$ for $1\le j\le M$.
\end{Remark}\vspace{-.3in}

\subsection{Aggregate loss and stop-loss moments}\label{subsec:RiskAggreStopLoss}\vspace{-.2in}

For a portfolio of dependent risks $\bm{X}=(X_1,\dots, X_M)$ with density $f\in\mathrm{MMEam}$, we are interested in the distribution of the aggregate loss $S=\sum_{j=1}^MX_j$. Like some other works in the literature (see e.g. \cite{willmot2015some} for multivariate mixed Erlang), it can be desirable if $S$ belongs to the same class of distributions as the individual risks so that calculations concerning the individual risks carry over to $S$ with a change of parameters. Denoting the density of $S$ by $f_S$, it can be shown that $f_S\in\mathrm{MEam}$ as follows.
\begin{theorem}\label{th:RiskAggregation} (\textbf{Aggregating components of $\mathrm{MMEam}$}) If the multivariate loss $(X_1, \dots, X_M)$ follows the density $f\in\mathrm{MMEam}$ of the form (\ref{eq:densityMMEm1}), then the density $f_S$ of the aggregate loss $S=\sum_{j=1}^MX_j$ belongs to $\mathrm{MEam}$ and is given by
\begin{equation}\label{eq:fSresult}
f_S(y) = \sum_{\bm{i}\in\mathscr{S}} p_{\bm{i}} \bm{\alpha}_{\bm{i}}e^{\bm{T}_{\bm{i}}y}\bm{t}_{\bm{i}}
\end{equation}
for $y\ge0$, where
\begin{equation}\label{eq:fScomponent}
\bm{\alpha}_{\bm{i}} = (\bm{\alpha}_{i_1},\bm{0},\bm{0},\dots,\bm{0}),\quad \bm{T}_{\bm{i}}=\begin{pmatrix}\bm{T}_{i_1} & \bm{t}_{i_1} \bm{\alpha}_{i_2}&&\\&\bm{T}_{i_2} & \bm{t}_{i_2} \bm{\alpha}_{i_3}&\\&&\ddots & \\&&&\bm{T}_{i_M}\end{pmatrix},\quad \bm{t}_{\bm{i}}=\begin{pmatrix}\bm{0}\\\bm{0}\\\vdots\\\bm{t}_{i_M}\end{pmatrix}.
\end{equation}
\end{theorem}\vspace{-.2in}
\begin{proof}
Using the joint density (\ref{eq:densityMMEm1}) of $(X_1,\dots, X_M)$, it is clear that
\[
f_S(y) = \sum_{\bm{i}\in\mathscr{S}} p_{\bm{i}} (f_{i_1} \ast f_{i_2} \ast \cdots \ast f_{i_M})(y).
\]
From Proposition \ref{prop:convolutionME1}, we observe that $f_{i_1} \ast f_{i_2} \ast \cdots \ast f_{i_M}$ is an $\mathrm{ME}$ density with parameters $(\bm{\alpha}_{\bm{i}},\bm{T}_{\bm{i}}, \bm{t}_{\bm{i}})$ given by (\ref{eq:fScomponent}), and the result (\ref{eq:fSresult}) follows. Clearly,  (\ref{eq:fSresult}) belongs to $\mathrm{MEam}$ according to the definition (\ref{eq:MEam1}).
\end{proof}\vspace{-.2in}

When a deductible $d\ge0$ is applied to the aggregate loss $S$ (instead of an individual risk), which is common in a \emph{stop-loss reinsurance} contract, then the ceded loss equals $(S-d)_+$. For $r\in\mathds{N}_+$, the $r$-th \emph{stop-loss moment} is defined by $\mathds{E}[(S-d)_+^r]$. It is important to be cautious about various actuarial terminologies in connection to Section \ref{subsec:ELandMCTE}. Excess-of-loss (re)insurance usually refers to the situation where a deductible is applied to an individual risk. On the other hand, in stop-loss (re)insurance and in the calculation of stop-loss moments, the deductible is applied to the aggregate loss. However, the terminology `stop-loss transform' can be used for a general random variable regardless of whether it is an individual risk or the aggregate loss. The following theorem presents the stop-loss moments of $S$.
\begin{theorem}\label{th:StopLoss} (\textbf{Stop-loss moments of aggregate loss from $\mathrm{MMEam}$}) Suppose that the multivariate loss $(X_1, \dots, X_M)$ follows the density $f\in\mathrm{MMEam}$ of the form (\ref{eq:densityMMEm1}). Then, for $r\in\mathds{N}_+$ and $d\ge0$, the $r$-th stop-loss moment of $S=\sum_{j=1}^MX_j$ is
\begin{equation}\label{eq:StopLossResult}
\mathds{E}[(S-d)_+^r] = \left(\sum_{\bm{i}\in\mathscr{S}} p^\mathrm{RL}_{S,d,\bm{i}} \left( r!\, \bm{\alpha}^\mathrm{RL}_{S,d,\bm{i}} (-\bm{T}_{\bm{i}})^{-(r+1)}\bm{t}_{\bm{i}} \right)\right) \left(\sum_{\bm{i}\in\mathscr{S}} p_{\bm{i}} \bm{\alpha}_{\bm{i}}e^{\bm{T}_{\bm{i}}d}\bm{l}_{\bm{i}}\right),
\end{equation}
where
\begin{equation}\label{eq:StopLossComponents}
p^\mathrm{RL}_{S,d,\bm{i}} =\frac{p_{\bm{i}}\left(\bm{\alpha}_{\bm{i}}e^{\bm{T}_{\bm{i}} d} \bm{l}_{\bm{i}}\right)}{\sum_{\bm{h}\in\mathscr{S}}p_{\bm{h}}\left(\bm{\alpha}_{\bm{h}}e^{\bm{T}_{\bm{h}} d} \bm{l}_{\bm{h}}\right)},\quad \bm{\alpha}^\mathrm{RL}_{S,d,\bm{i}} = \frac{1}{\bm{\alpha}_{\bm{i}}e^{\bm{T}_{\bm{i}} d} \bm{l}_{\bm{i}}} \bm{\alpha}_{\bm{i}}e^{\bm{T}_{\bm{i}} d}
\end{equation}
with the definition $\bm{l}_{\bm{i}}=(-\bm{T}_{\bm{i}})^{-1}\bm{t}_{\bm{i}}$ for $\bm{i}\in\mathscr{S}$. 
\end{theorem}\vspace{-.2in}
\begin{proof}
Since $f_S\in\mathrm{MEam}$ has representation (\ref{eq:fSresult}) according to Theorem \ref{th:RiskAggregation}, application of the univariate version of Theorem \ref{th:residual1} reveals that the residual life time variable $S-d|S>d$ has the $\mathrm{MEam}$ density
\begin{equation}\label{eq:fSRL}
f_{S,d}^\mathrm{RL}(y) = \sum_{\bm{i}\in\mathscr{S}} p^\mathrm{RL}_{S,d,\bm{i}} \left(\bm{\alpha}^\mathrm{RL}_{S,d,\bm{i}}e^{\bm{T}_{\bm{i}} y} \bm{t}_{\bm{i}}\right)
\end{equation}
for $y\ge0$ with components given by (\ref{eq:StopLossComponents}). From the univariate version of Theorem \ref{th:crossmomentsRL} (or by direct integration using (\ref{eq:fSRL})), one has
\begin{equation}\label{eq:StopLossConditional}
\mathds{E}[(S-d)^r|S>d] =  \sum_{\bm{i}\in\mathscr{S}} p^\mathrm{RL}_{S,d,\bm{i}} \left( r!\, \bm{\alpha}^\mathrm{RL}_{S,d,\bm{i}} (-\bm{T}_{\bm{i}})^{-(r+1)}\bm{t}_{\bm{i}} \right).
\end{equation}
Multiplication by the survival function of $S$, namely $\overline{F}_S(d) = \sum_{\bm{i}\in\mathscr{S}} p_{\bm{i}} \bm{\alpha}_{\bm{i}}e^{\bm{T}_{\bm{i}}d}\bm{l}_{\bm{i}}$, yields the desired result (\ref{eq:StopLossResult}).
\end{proof}\vspace{-.3in}

\subsection{Largest claims and ECOMOR reinsurance treaties}\label{subsec:LargeClaimsRein}\vspace{-.2in}

In addition to excess-of-loss and stop-loss reinsurance contracts discussed in the previous subsections, our results on order statistics can also be applied to calculate the pure premium of reinsurance treaties based on the large claims in the portfolio (e.g. \cite{Kremer1982LCR, Kremer1998LCR, Berglund1998LCR, LadoucetteTeugels2006LCR}). Recall that $X_{j: M}^\mathrm{OS}$ is the $j$-th order statistic of the $M$ dependent losses $\bm{X}=(X_1,\dots, X_M)$, arranged from the smallest to the largest. Letting $g_j$ be the ceded loss function for the $j$-th order statistic satisfying $0\le g_j(x) \le x$ for $x\ge0$ and $1\le j\le M$, the total loss covered by the reinsurer is generally given by $R=\sum_{j=1}^M g_j(X_{j: M}^\mathrm{OS})$. Therefore, the pure reinsurance premium is simply
\begin{equation}\label{eq:ReinPrem}
\mathds{E}[R] = \sum_{j=1}^M \mathds{E}[g_j(X_{j: M}^\mathrm{OS})].
\end{equation}
In general, the functions $g_j$'s do not have to be of the same form. For example, if $g_j(x)=a_jx$ for some $0\le a_j\le 1$, then the loss $X_{j: M}^\mathrm{OS}$ is subject to proportional reinsurance and $\mathds{E}[g_j(X_{j: M}^\mathrm{OS})]=a_j\mathds{E}[X_{j: M}^\mathrm{OS}]$. On the other hand, if $g_j(x)=(x-z_j)_+$ for some $z_j\ge0$, then $X_{j: M}^\mathrm{OS}$ is subject to excess-of-loss reinsurance and $\mathds{E}[g_j(X_{j: M}^\mathrm{OS})]=\mathds{E}[(X_{j: M}^\mathrm{OS}-z_j)_+]$. With $f\in\mathrm{MMEam}$, it is known from Theorem \ref{th:OS} that $X_{j: M}^\mathrm{OS}$ follows the $\mathrm{MEam}$ density (\ref{eq:orderm1}) with $f_{\bm{i},j:M}$ an $\mathrm{ME}$ density. Denoting the triple of $f_{\bm{i},j:M}$ by $(\bm{\alpha}_{\bm{i},j:M},\bm{T}_{\bm{i},j:M},\bm{t}_{\bm{i},j:M})$ and defining $\bm{l}_{\bm{i},j:M}=(-\bm{T}_{\bm{i},j:M})^{-1}\bm{t}_{\bm{i},j:M}$, the afore-mentioned examples of expectation are given by 
\begin{equation}\label{eq:OSmean}
\mathds{E}[X_{j: M}^\mathrm{OS}] =\sum_{\bm{i}\in\mathscr{S}}p_{\bm{i}} \left(\bm{\alpha}_{\bm{i},j:M}(-\bm{T}_{\bm{i},j:M})^{-2}\bm{t}_{\bm{i},j:M}\right)
\end{equation}
and
\[
\mathds{E}[(X_{j: M}^\mathrm{OS}-z_j)_+] =\left(\sum_{\bm{i}\in\mathscr{S}}p^\mathrm{RL}_{z_j,\bm{i},j:M} \left(\bm{\alpha}^\mathrm{RL}_{z_j,\bm{i},j:M} (-\bm{T}_{\bm{i},j:M})^{-2}\bm{t}_{\bm{i},j:M}\right)\right) \left(\sum_{\bm{i}\in\mathscr{S}} p_{\bm{i}} \bm{\alpha}_{\bm{i},j:M}e^{\bm{T}_{\bm{i},j:M}z_i}\bm{l}_{\bm{i},j:M}\right),
\]
where
\[
p^\mathrm{RL}_{z_j,\bm{i},j:M} =\frac{p_{\bm{i}}\left(\bm{\alpha}_{\bm{i},j:M}e^{\bm{T}_{\bm{i},j:M} z_j} \bm{l}_{\bm{i},j:M}\right)}{\sum_{\bm{h}\in\mathscr{S}}p_{\bm{h}}\left(\bm{\alpha}_{\bm{h},j:M}e^{\bm{T}_{\bm{h},j:M} z_j} \bm{l}_{\bm{h},j:M}\right)},\quad \bm{\alpha}^\mathrm{RL}_{z_j,\bm{i},j:M}  = \frac{1}{\bm{\alpha}_{\bm{i},j:M}e^{\bm{T}_{\bm{i},j:M} z_j} \bm{l}_{\bm{i},j:M}} \bm{\alpha}_{\bm{i},j:M}e^{\bm{T}_{\bm{i},j:M} z_j}.
\]

In the usual $k$ \emph{largest claims reinsurance} ($\mathrm{LCR}(k)$) for $\bm{X}=(X_1,\dots, X_M)$ where $1\le k\le M$, the reinsurer fully covers the largest $k$ losses so that $R=\sum_{j=M-k+1}^MX_{j: M}^\mathrm{OS}$. Therefore, the pure reinsurance premium (\ref{eq:ReinPrem}) can be calculated by defining $g_j$ as $g_j(x)=x\mathds{1}\{M-k+1\le j\le M\}$ and using (\ref{eq:OSmean}). Another interesting reinsurance contract is the ECOMOR($k$) (\emph{Exc\'edent du Co\^ut Moyen Relatif}) treaty with $2\le k\le M$, where the reinsurer covers the losses in excess of the $k$-th largest loss such that
\[
R=\sum_{j=1}^M(X_{j: M}^\mathrm{OS}-X_{M-k+1: M}^\mathrm{OS})_+ = \sum_{j=M-k+2}^M(X_{j: M}^\mathrm{OS}-X_{M-k+1: M}^\mathrm{OS}) = \sum_{j=M-k+2}^M X_{j: M}^\mathrm{OS} - (k-1) X_{M-k+1: M}^\mathrm{OS}.
\]
ECOMOR($k$) is like excess-of-loss reinsurance for the $k-1$ largest losses, where the deductible is random and is taken to be the $k$-th largest loss. From the final expression above, $R$ is retrieved by letting $g_j(x)=x\mathds{1}\{M-k+2\le j\le M\}-(k-1)x\mathds{1}\{j=M-k+1\}$, and therefore $\mathds{E}[R]$ can be evaluated using (\ref{eq:OSmean}).
\vspace{-.2in}

\subsection{Conditional Value-at-Risk ($\mathrm{CoV@R}$)}\label{subsec:RiskMeasures}\vspace{-.2in}

The concept of \emph{conditional Value-at-Risk}, abbreviated as $\mathrm{CoV@R}$, was first introduced by \cite{adrian2016CoVaR}. The motivation of $\mathrm{CoV@R}$ can be better explained by interpreting the components of $\bm{X}=(X_1,\dots, X_M)$ as the losses faced by different lines of business. Then, $\mathrm{CoV@R}$ can be defined as the $\mathrm{V@R}$ of a loss or the total loss conditional on a stress scenario (e.g. another business line is in financial distress). For example, following \cite{adrian2016CoVaR}, the risk measure $\mathrm{CoV@R}_{\theta_1,\theta_2}^=(X_2|X_1)=\mathrm{V@R}_{\theta_2}(X_2|X_1=\mathrm{V@R}_{\theta_1}(X_1))$ is the $\theta_2$-level $\mathrm{V@R}$ of $X_2$ given that the loss $X_1$ is exactly at its $\theta_1$-level $\mathrm{V@R}$, where $\theta_1,\theta_2\in[0,1)$. \cite{Girardi2013CoVaR} subsequently modified the condition and defined $\mathrm{CoV@R}_{\theta_1,\theta_2}^>(X_2|X_1)=\mathrm{V@R}_{\theta_2}(X_2|X_1>\mathrm{V@R}_{\theta_1}(X_1))$ to include severe losses beyond line 1's $\mathrm{V@R}$ in the stress scenario. In particular, \cite{Mainik2014CoVaR} suggested that $\mathrm{CoV@R}_{\theta_1,\theta_2}^>(X_2|X_1)$ could be a better risk measure than $\mathrm{CoV@R}_{\theta_1,\theta_2}^=(X_2|X_1)$ as far as dependence consistency is concerned.

For either definition of $\mathrm{CoV@R}$, when $\bm{X}=(X_1,\dots, X_M)$ follows an $\mathrm{MMEam}$ distribution, the closure properties of the $\mathrm{MMEam}$ class under conditional distributions in Section \ref{sec:ConPDF} and under residual lifetime distributions in Section \ref{sec:RL} make it convenient for us to obtain the relevant conditional cumulative distributions so as to compute $\mathrm{CoV@R}$. Specifically, for $x_1,x_2\ge0$ the conditional cumulative distribution function of $X_2|X_1=x_1$ is
\[
F_{X_2|X_1}^=(x_2|x_1) = 1- \sum_{\bm{i}\in\mathscr{S}} p^\mathrm{con}_{x_1,\bm{i}}  \bm{\alpha}_{i_2}e^{\bm{T}_{i_2}x_2}\bm{l}_{i_2},\quad p^\mathrm{con}_{x_1,\bm{i}} = \frac{p_{\bm{i}} \bm{\alpha}_{i_1}e^{\bm{T}_{i_1}x_1}\bm{t}_{i_1}}{\sum_{\bm{h}\in\mathscr{S}}p_{\bm{h}} \bm{\alpha}_{h_1}e^{\bm{T}_{h_1}x_1}\bm{t}_{h_1}},
\]
and that of $X_2|X_1>x_1$ is
\[
F_{X_2|X_1}^>(x_2|x_1) = 1- \sum_{\bm{i}\in\mathscr{S}} p^\mathrm{RL}_{x_1,0,\dots,0,\bm{i}} \bm{\alpha}_{i_2}e^{\bm{T}_{i_2}x_2}\bm{l}_{i_2},\quad p^\mathrm{RL}_{x_1,0,\dots,0,\bm{i}} = \frac{p_{\bm{i}}\bm{\alpha}_{i_1}e^{\bm{T}_{i_1} x_1} \bm{l}_{i_1}}{\sum_{\bm{h}\in\mathscr{S}}p_{\bm{h}}\bm{\alpha}_{h_1}e^{\bm{T}_{h_1} x_1} \bm{l}_{h_1}},
\]
where the notation $p^\mathrm{RL}_{x_1,0,\dots,0,\bm{i}}$ follows (\ref{eq:residualcom2MMem1}) by taking $\bm{z}=(x_1,0,\dots,0)$. Thus, $\mathrm{CoV@R}_{\theta_1,\theta_2}^=(X_2|X_1)$ satisfies the equation $F_{X_2|X_1}^=(\mathrm{CoV@R}_{\theta_1,\theta_2}^=(X_2|X_1)|\mathrm{V@R}_{\theta_1}(X_1)) = \theta_2$ whereas $\mathrm{CoV@R}_{\theta_1,\theta_2}^>(X_2|X_1)$ can be solved from $F_{X_2|X_1}^>(\mathrm{CoV@R}_{\theta_1,\theta_2}^>(X_2|X_1)|\mathrm{V@R}_{\theta_1}(X_1)) = \theta_2$. We remark that \cite{Mainik2014CoVaR} also defined \emph{conditional expected shortfall} ($\mathrm{CoES}$), which is an analogue of $\mathrm{ES}$ (same as $\mathrm{TCE}$ for continuous distribution) that is conditional on a stress scenario. This can be easily computed using the above conditional distributions and the details are omitted. Moreover, the conditions in the stress scenario can be generalized such that multiple lines have losses at or exceeding their respective $\mathrm{V@R}$ values, and the relevant $\mathrm{CoV@R}$ and $\mathrm{CoES}$ can still be calculated because of the closure properties.

Referring to e.g. \cite{DiBernardino2015MultiCoVaR} for review and further discussions of $\mathrm{CoV@R}$, we note that an interesting $\mathrm{CoV@R}$ relevant to insurance can be defined for the aggregate loss $S=\sum_{j=1}^MX_j$ conditional on a specific loss, say $X_1$, being large. We shall consider $\mathrm{CoV@R}_{\theta_1,\theta_S}^>(S|X_1)=\mathrm{V@R}_{\theta_S}(S|X_1>\mathrm{V@R}_{\theta_1}(X_1))$ as follows. To determine the conditional distribution $S|X_1>z_1$ for $z_1\ge0$, we first write $S=z_1 + (X_1-z_1)+\sum_{j=2}^MX_j$. Setting $z_2=\dots=z_m=0$ and $z_1\ge0$ in Theorem \ref{th:residual1}, with $(X_1,\dots, X_M)$ an $\mathrm{MMEam}$ distribution the joint density of $(X_1-z_1,X_2,\dots, X_M)$ given $X_1>z_1$ is found to be
\[
f^\mathrm{RL}_{z_1,0,\dots,0}(x_1,\dots,x_M) = \sum_{\bm{i}\in\mathscr{S}} p^\mathrm{RL}_{z_1,0,\dots,0,\bm{i}}\,f^\mathrm{RL}_{z_1,i_1}(x_1) f_{i_2}(x_2) \cdots f_{i_M}(x_M)
\]
for $x_1, \dots , x_M\ge 0$, and $f^\mathrm{RL}_{z_1,i_1}\in\mathrm{ME}$ is defined via (\ref{eq:residualcom1MMem1}). One can then easily follow Theorem \ref{th:RiskAggregation} to see that the sum $(X_1-z_1)+\sum_{j=2}^MX_j$ (which equals $S-z_1$) has conditional cumulative distribution function
\[
F_{S-z_1|X_1}^>(y|z_1) = \sum_{\bm{i}\in\mathscr{S}}  p^\mathrm{RL}_{z_1,0,\dots,0,\bm{i}} \widetilde{\bm{\alpha}}_{z_1,\bm{i}} e^{\bm{T}_{\bm{i}}y}\bm{t}_{\bm{i}}
\]
for $y\ge0$, where $\bm{T}_{\bm{i}}$ and $\bm{t}_{\bm{i}}$ has the same definitions as in (\ref{eq:fScomponent}) and 
$\widetilde{\bm{\alpha}}_{z_1,\bm{i}} = (\bm{\alpha}^\mathrm{RL}_{z_1,i_1},\bm{0},\bm{0},\dots,\bm{0})$. Taking $z_1=\mathrm{V@R}_{\theta_1}(X_1)$ and using the translation invariance of $\mathrm{V@R}$ (and hence $\mathrm{CoV@R}$), we have $\mathrm{CoV@R}_{\theta_1,\theta_S}^>(S|X_1) = \mathrm{V@R}_{\theta_1}(X_1) + \mathrm{CoV@R}_{\theta_1,\theta_S}^>(S-\mathrm{V@R}_{\theta_1}(X_1)|X_1)$, where $\mathrm{CoV@R}_{\theta_1,\theta_S}^>(S-\mathrm{V@R}_{\theta_1}(X_1)|X_1)$ satisfies $F_{S-\mathrm{V@R}_{\theta_1}(X_1)|X_1}^>(\mathrm{CoV@R}_{\theta_1,\theta_S}^>(S-\mathrm{V@R}_{\theta_1}(X_1)|X_1)|\mathrm{V@R}_{\theta_1}(X_1))=\theta_S$.
\vspace{-.2in}

\subsection{Weighted premium calculations}\label{subsec:WeightedPC}\vspace{-.2in}

In this subsection, we shall discuss how our results in Section \ref{subsec:SBET} concerning size-biased Esscher transform of an $\mathrm{MMEam}$ distribution are connected to \emph{weighted premium calculations}, a concept introduced in \cite{FurmanZitikis2008weightPC}. For the vector of losses $\bm{X}=(X_1, \dots, X_M)$ with density $f$, the weighted version of the vector of variables is denoted by $\bm{X}^{\{w\}}=(X_1^{\{w\}}, \dots, X_M^{\{w\}})$ and it has joint density
\begin{equation}\label{eq:weightPDFDef}
f^{\{w\}}(x_1,\dots,x_M) = \frac{w(x_1,\dots,x_M)}{\mathds{E}[w(X_1, \dots, X_M)]} f(x_1,\dots,x_M)
\end{equation}
for $x_1,\dots,x_M\ge 0$, where the weight $w$ is a non-negative function on $[0,\infty)^M$ such that $\mathds{E}[w(\bm{X})]$ is positive and finite. See e.g. \cite{navarro2006MultiSizeBiased} for a review of multivariate weighted distributions. In the univariate case (i.e. $M=1$), \cite{FurmanZitikis2008weightPC} defined the weighted premium for the risk $X_1$ as $\mathds{E}[X_1w(X_1)]/\mathds{E}[w(X_1)]=\mathds{E}[X_1^{\{w\}}]$, which is the expectation of the weighted random variable $X_1^{\{w\}}$. It is well known in the univariate case that if $w$ is increasing (i.e. non-decreasing) then $X_1$ is smaller than $X_1^{\{w\}}$ in likelihood ratio order (denoted by $X_1 \le _{\mathrm{lr}} X_1^{\{w\}}$), which implies the stochastic order $X_1 \le _{\mathrm{st}} X_1^{\{w\}}$. See e.g. \cite[Chapter 1.C.1]{ShakedShanthikumar2007Book}. Consequently, the weighted premium has a non-negative loading, i.e. $\mathds{E}[X_1^{\{w\}}]\ge \mathds{E}[X_1]$, when $w$ is increasing. Extensions of the univariate weighted premium calculations of the risk $X_1$ was made by \cite{FurmanZitikis2008weightCA} such that the weight is allowed to be a function of another random variable $X_2$. This corresponds to $M=2$ in the general formulation with $w(x_1,x_2) = w_2(x_2)$ depending on the second argument only so that the weighted premium is $\mathds{E}[X_1^{\{w_2\}}]=\mathds{E}[X_1 w_2(X_2)]/\mathds{E}[w_2(X_2)]$. In general, $X_1$ and $X_2$ are assumed dependent, otherwise $\mathds{E}[X_1^{\{w_2\}}]$ simply reduces to $\mathds{E}[X_1]$. While the focus of \cite{FurmanZitikis2008weightCA} is on capital allocation, the calculations were further generalized by \cite{Zhu2015thesis, Zhu2019MultiWeighted} to the multivariate case where the weighted premium for a given risk, say $X_1$, is $\mathds{E}[X_1^{\{w\}}]=\mathds{E}[X_1w(\bm{X})]/\mathds{E}[w(\bm{X})]$. Note that one needs the condition $\mathrm{Cov}(X_1,w(\bm{X}))\ge 0$ to ensure the premium is loaded.

With $\bm{X}=(X_1, \dots, X_M)$ assumed to follow an $\mathrm{MMEam}$ density of the form (\ref{eq:densityMMEm1}), its size-biased Esscher transformed density $f^\mathrm{ET}_{\bm{n},\bm{\lambda}}$ in (\ref{eq:densityGenEsscher}) is a weighted density with weight function
\begin{equation}\label{eq:weightSBET}
 w(x_1,\dots,x_M) = x_1^{n_1}\dots x_M^{n_M} e^{-\lambda_1 x_1 \cdots - \lambda_M x_M},
\end{equation}
where $n_i\in\mathds{N}_0$ and $\lambda_i>\max_{1\le j \le L}\kappa_j$ for $1\le i\le M$ (see Remark \ref{rem:Esschertransform}). More importantly, by Theorem \ref{th:SBET} the density $f^\mathrm{ET}_{\bm{n},\bm{\lambda}}$ still belongs to $\mathrm{MMEam}$ which can make calculations convenient. Note that if $\max_{1\le j \le L}\kappa_j<\lambda_i\le 0$, then $w$ in \eqref{eq:weightSBET} is an increasing function. In this case, if the vector $\bm{X}$ is associated, then from \cite[Theorem 6.B.8]{ShakedShanthikumar2007Book} one has the multivariate stochastic order $\bm{X} \le _{\mathrm{st}} \bm{X}^{\{w\}}$, i.e. $\mathds{E}[a(\bm{X})]\le \mathds{E}[a(\bm{X}^{\{w\}})]$ for all increasing functions $a$ where the expectations exist. (A vector $\bm{X}$ is said to be associated if $\mathrm{Cov}(a_1(\bm{X}),a_2(\bm{X}))\ge0$ for all increasing functions $a_1$ and $a_2$ such that the covariance is defined \cite{Esary1967Association}.) The property $\mathds{E}[a(\bm{X})]\le \mathds{E}[a(\bm{X}^{\{w\}})]$ for increasing $a$ can be desirable for premium calculations even when (some of) the variables contained in $\bm{X}=(X_1, \dots, X_M)$ are not the losses but are instead risk factors contributing to the losses. For example, the losses arising from a storm can depend on factors including the scale of the storm, the area affected by the storm and the storm's duration in a multiplicative manner, and the total loss may be represented as $a(\bm{X})$ with $a$ involving additive and multiplicative components. The inequality $\mathds{E}[a(\bm{X})]\le \mathds{E}[a(\bm{X}^{\{w\}})]$ ensures that the weighted premium is loaded.

It is worthwhile to point out that the (first order) multivariate equilibrium distribution can also be regarded as a weighted distribution \cite[Section 3]{navarro2006MultiSizeBiased}. Specifically, (\ref{eq:equilibriumdistGen}) at $r=1$ can be rewritten in the form of (\ref{eq:weightPDFDef}) where $w(x_1,\dots,x_M)=\overline{F}(x_1,\dots,x_M)/f(x_1,\dots,x_M) = 1/\mu(x_1,\dots,x_M)$ is the reciprocal of the multivariate failure rate function $\mu$ for the vector $\bm{X}=(X_1, \dots, X_M)$ (see \cite{Puri1974MultiFailureRate}). Therefore, if $\mu$ is a decreasing (i.e. non-increasing) function then $w$ is an increasing function. Again, if $\bm{X}$ is associated then $\bm{X} \le _{\mathrm{st}} \bm{X}^{\{w\}}$, and the same comments on weighted premium calculations as in the size-biased Esscher transform apply.
\vspace{-.2in}

\subsection{Risk capital allocation}\label{subsec:riskcapital1}\vspace{-.2in}

In capital allocation, a risk measure $\rho(S)$ of the aggregate risk $S=\sum_{j=1}^MX_j$ can be regarded as the \emph{risk capital} of the firm, and the objective is to allocate the risk capital to each risk according to an \emph{allocation principle}. A reasonable allocation principle should possess certain properties, and we refer interested readers to \cite{Denault2001RAC} for the details. Denoting $K_j$ the capital allocated to the risk $X_j$ for $1\le j\le M$, it is often desirable for an allocation principle to satisfy the additivity requirement $\rho(S)=\sum_{j=1}^M K_j$. Our focus here is to show that the intermediate quantities appearing in various allocation principles can be readily evaluated when the vector of dependent risks $\bm{X}=(X_1,\dots, X_M)$ has density $f\in\mathrm{MMEam}$.
\begin{Example}\label{ex:CapitalCov} (\textbf{Covariance-based allocation for $\mathrm{MMEam}$ risks}) Let us consider an allocation principle in relation to the covariance between a given risk $X_j$ and the aggregate risk $S$. The risk measure $\rho$ is taken to be the $\theta$-level $\mathrm{TV@R}$ (which is the same as $\mathrm{TCE}$ since $S$ is continuous). From e.g. \cite[Equation (10)]{Cossette2013FGMErlang}, for $1\le j\le M$ the allocated capital $K_j$ to the $j$-th risk according to a covariance-based allocation is
\[
C_\theta(X_j,S) = \mathds{E}[X_j] + \frac{\mathrm{Cov}(X_j,S)}{\mathrm{Var}(S)} (\mathrm{TV@R}_\theta(S)-\mathds{E}[S]).
\]
Clearly, $\sum_{j=1}^M C_\theta(X_j,S)=\mathrm{TV@R}_\theta(S)$ so that additivity is satisfied. With $f\in\mathrm{MMEam}$, we can find all components in the above allocation rule. First, $\mathds{E}[S]$ and $\mathds{E}[S^2]$ (and hence $\mathrm{Var}(S)=\mathds{E}[S^2]-(\mathds{E}[S])^2$) follow from (\ref{eq:StopLossConditional}) by letting $d=0$ and $r=1,2$. Second, the covariance $\mathrm{Cov}(X_j,S)$ can be written as $\mathrm{Cov}(X_j,S) = \mathds{E}[X_jS]-\mathds{E}[X_j]\mathds{E}[S] = \sum_{i=1}^M\mathds{E}[X_jX_i] -\mathds{E}[X_j]\mathds{E}[S]$, where the moments $\mathds{E}[X_j]$, $\mathds{E}[X_j^2]$ and $\mathds{E}[X_jX_i]$ (for $i\ne j$) are available from Theorem \ref{th:crossmoments1}. Third, from (\ref{eq:fSresult}) it is clear that the cumulative distribution function of $S$ is $F_S(y) = 1- \sum_{\bm{i}\in\mathscr{S}} p_{\bm{i}} \bm{\alpha}_{\bm{i}}e^{\bm{T}_{\bm{i}}y}\bm{l}_{\bm{i}}$ for $y\ge0$, and therefore $\mathrm{V@R}_\theta(S)$ can be solved from $F_S(\mathrm{V@R}_\theta(S))=\theta$, and then one can utilize (\ref{eq:StopLossConditional}) with $r=1$ to compute
\[
\mathrm{TV@R}_\theta(S) = \mathrm{V@R}_\theta(S) + \sum_{\bm{i}\in\mathscr{S}} p^\mathrm{RL}_{S,\mathrm{V@R}_\theta(S),\bm{i}} \left(\bm{\alpha}^\mathrm{RL}_{S,\mathrm{V@R}_\theta(S),\bm{i}} (-\bm{T}_{\bm{i}})^{-2}\bm{t}_{\bm{i}} \right).
\]
\end{Example}\vspace{-.2in}

In Theorem \ref{th:JointTailMomentsXS}, we shall provide closed-form formulas for the quantity $\mathds{E}[X^k_jS^h\mathds{1}\{S>y\}]$ for $k,h\in\mathds{N}_0$, which will be useful for other capital allocation rules to be discussed in Examples \ref{ex:TCovRCA} and \ref{ex:TCPA}. The proof of Theorem \ref{th:JointTailMomentsXS} requires the following lemma.
\begin{Lemma}\label{lem:intmomentsyinfty} Let $\bm{B}$ be a square matrix whose eigenvalues have strictly negative real parts. Then, for $n\in\mathds{N}_0$ and $y\ge0$ we have
\[
\int_y^\infty u^n e^{\bm{B} u}\dd u = e^{\bm{B}y}\sum_{k=0}^n (-\bm{B})^{-(n-k+1)}\frac{n!}{k!} y^k. 
\]
\end{Lemma}\vspace{-.2in}
\begin{proof}
The assumption on $\bm{B}$ implies that the norm of $u^n e^{\bm{B} u}$ decays exponentially as $u\rightarrow\infty$, and thus $\int_y^\infty u^n e^{\bm{B} u}\dd u$ is finite. Letting $\bm{Q}(y)=e^{\bm{B}y}\sum_{k=0}^n (-\bm{B})^{-(n-k+1)}\tfrac{n!}{k!} y^k$, we take derivative to arrive at
\begin{align*}
\frac{\dd \bm{Q}(y)}{\dd y} & = \left(\bm{B}e^{\bm{B}y}\sum_{k=0}^n (-\bm{B})^{-(n-k+1)}\frac{n!}{k!} y^k\right) + \left(e^{\bm{B}y}\sum_{k=1}^n (-\bm{B})^{-(n-k+1)}\frac{n!}{k!} k y^{k-1} \right)\\
& = \left(-e^{\bm{B}y}\sum_{k=1}^{n+1} (-\bm{B})^{-(n-k+1)}\frac{n!}{(k-1)!} y^{k-1}\right) + \left(e^{\bm{B}y}\sum_{k=1}^n (-\bm{B})^{-(n-k+1)}\frac{n!}{(k-1)!} y^{k-1} \right)\\
& = -e^{\bm{B} y} y^n,
\end{align*}
and the proof is complete.
\end{proof}\vspace{-.2in}

Since the results in the following theorem will be used for capital allocation, it suffices to assume that $M\ge 2$, i.e. there are at least two risks to which the capital is allocated.
\begin{theorem}\label{th:JointTailMomentsXS} (\textbf{Cross moments of individual and total risks under $\mathrm{MMEam}$}) Suppose that $(X_1, \dots, X_M)$ follows the density $f\in\mathrm{MMEam}$ of the form (\ref{eq:densityMMEm1}). Fix $M\ge 2$ and $j\in\{1,\dots,M\}$. For $h\in\mathds{N}_0$ and $y\ge 0$, the cross moments $\mathds{E}[X^k_jS^h\mathds{1}\{S>y\}]$ of $X_j$ and $S=\sum_{i=1}^MX_i$ on the set $\{S>y\}$ are given by 
\begin{equation}\label{eq:MomentsXSk0}
\mathds{E}[S^h\mathds{1}\{S>y\}] = \sum_{\bm{i}\in\mathscr{S}} p_{\bm{i}} \bm{\alpha}_{\bm{i}} \left(e^{\bm{T}_{\bm{i}}y}\sum_{\ell=0}^h (-\bm{T}_{\bm{i}})^{-(h-\ell+1)}\frac{h!}{\ell!} y^\ell\right) \bm{t}_{\bm{i}}
\end{equation}
when $k=0$, and by
\begin{equation}\label{eq:capitalaux1}
\mathds{E}[X^k_jS^h\mathds{1}\{S>y\}] = \sum_{\bm{i}\in\mathscr{S}} p_{\bm{i}}  k! \left(\bm{\alpha}_{i_j}^{[k]}, \bm{0}\right)\left\{e^{\bm{A}^{\{k,\bm{i},j\}}y}\sum_{\ell=0}^h \left(-\bm{A}^{\{k,\bm{i},j\}}\right)^{-(h-\ell+1)}\frac{h!}{\ell!} y^\ell\right\}\begin{pmatrix}\bm{0}\\ \bm{t}_{(\bm{i},j)}\end{pmatrix}
\end{equation}
when $k\in\mathds{N}_+$. Here we have defined
\begin{equation}\label{eq:Acapitalaux1}
\bm{A}^{\{k,\bm{i},j\}}=\begin{pmatrix}\bm{T}_{i_j}^{[k,0]}& \bm{t}_{i_j}^{[k]} \bm{\alpha}_{(\bm{i},j)}\\ \bm{0}& \bm{T}_{(\bm{i},j)}\end{pmatrix},
\end{equation}
and $(\bm{\alpha}_{(\bm{i},j)}, \bm{T}_{(\bm{i},j)}, \bm{t}_{(\bm{i},j)})$ are the parameters of $f_{i_1} \ast f_{i_2} \ast \cdots \ast f_{i_{j-1}}\ast f_{i_{j+1}} \ast \cdots\ast f_{i_M}$ which belongs to $\mathrm{ME}$ as in Proposition \ref{prop:convolutionME1}.
\end{theorem}\vspace{-.2in}
\begin{proof}
We first start with the case $k=0$, so that $\mathds{E}[S^h\mathds{1}\{S>y\}]$ can be directly obtained by integrating the density (\ref{eq:fSresult}) to yield
\[
\mathds{E}[S^h\mathds{1}\{S>y\}] = \int_y^\infty  v^h\left(\sum_{\bm{i}\in\mathscr{S}} p_{\bm{i}} \bm{\alpha}_{\bm{i}}e^{\bm{T}_{\bm{i}}v}\bm{t}_{\bm{i}}\right) \dd v
\]
from which (\ref{eq:MomentsXSk0}) follows with the help of Lemma \ref{lem:intmomentsyinfty}.

Next, we consider $k\in\mathds{N}_+$. Define $S_{-j}=S-X_j$ as the total risk excluding the $j$-th risk. Let $f_{X_j,S}$ denote the bivariate density of $(X_j,S)$, and likewise let $f_{X_j,S_{-j}}$ be the bivariate density of $(X_j,S_{-j})$. Then, for $0\le x\le v$ one has
\begin{align*}
f_{X_j, S}(x,v) & = f_{X_j,S_{-j}}(x,v-x)\\
& = \sum_{\bm{i}\in\mathscr{S}} p_{\bm{i}} f_{i_j}(x) (f_{i_1} \ast f_{i_2} \ast \cdots \ast f_{i_{j-1}}\ast f_{i_{j+1}} \ast \cdots\ast f_{i_M})(v-x)\\
& = \sum_{\bm{i}\in\mathscr{S}} p_{\bm{i}} f_{i_j}(x) f^\mathrm{sum}_{(\bm{i},j)}(v-x),
\end{align*}
where $f^\mathrm{sum}_{(\bm{i},j)} = f_{i_1} \ast f_{i_2} \ast \cdots \ast f_{i_{j-1}}\ast f_{i_{j+1}} \ast \cdots\ast f_{i_M}$. Note that Proposition \ref{prop:convolutionME1} implies $f^\mathrm{sum}_{(\bm{i},j)}\in\mathrm{ME}$, say, with parameters $(\bm{\alpha}_{(\bm{i},j)}, \bm{T}_{(\bm{i},j)}, \bm{t}_{(\bm{i},j)})$. Using the above joint density, the cross moment of our concern is given by the integral
\begin{align}\label{eq:integralStep0}
\mathds{E}[X_j^kS^h\mathds{1}\{S>y\}] & =  \int_y^\infty v^h\int_0^v x^k f_{X_j, S}(x,v)\dd x\dd v\nonumber\\
& = \sum_{\bm{i}\in\mathscr{S}} p_{\bm{i}} \int_y^\infty v^h\int_0^v x^k f_{i_j}(x) f^\mathrm{sum}_{(\bm{i},j)}(v-x)\dd x\dd v.
\end{align}
The inner integral can be evaluated as
\begin{align*}
\int_0^v x^k f_{i_j}(x) f^\mathrm{sum}_{(\bm{i},j)}(v-x)\dd x& =\int_0^v \left(x^k \bm{\alpha}_{i_j}e^{\bm{T}_{i_j}x}\bm{t}_{i_j}\right)\left(\bm{\alpha}_{(\bm{i},j)}e^{\bm{T}_{(\bm{i},j)}(v-x)}\bm{t}_{(\bm{i},j)}\right) \dd x\nonumber\\
& = k! \int_0^v \left(\bm{\alpha}_{i_j}^{[k]}e^{\bm{T}_{i_j}^{[k,0]}x}\bm{t}_{i_j}^{[k]}\right)\left(\bm{\alpha}_{(\bm{i},j)}e^{\bm{T}_{(\bm{i},j)}(v-x)}\bm{t}_{(\bm{i},j)} \right)\dd x\nonumber\\
& = k! \left(\bm{\alpha}_{i_j}^{[k]}, \bm{0}\right) e^{\bm{A}^{\{k,\bm{i},j\}}v}\begin{pmatrix}\bm{0}\\ \bm{t}_{(\bm{i},j)}\end{pmatrix},
\end{align*}
where $\bm{A}^{\{k,\bm{i},j\}}$ is defined in (\ref{eq:Acapitalaux1}). In obtaining the above result, Lemma \ref{lem:momentdist1} and the notation in (\ref{eq:alphantn1}) have been used in the second equality and the matrix integration formula in \cite[Theorem 1]{van1978computing} in the last one. Then, substitution into (\ref{eq:integralStep0}) gives rise to
\[
\mathds{E}[X^k_jS^h\mathds{1}\{S>y\}] = \sum_{\bm{i}\in\mathscr{S}} p_{\bm{i}}  k! \left(\bm{\alpha}_{i_j}^{[k]}, \bm{0}\right)\left\{\int_y^\infty v^h e^{\bm{A}^{\{k,\bm{i},j\}}v} \dd v\right\}\begin{pmatrix}\bm{0}\\ \bm{t}_{(\bm{i},j)}\end{pmatrix},
\]
and the use of Lemma \ref{lem:intmomentsyinfty} leads us to (\ref{eq:capitalaux1}), completing the proof.
\end{proof}
\begin{Example}\label{ex:TCovRCA} (\textbf{$\mathrm{TCov}$ allocation for $\mathrm{MMEam}$ risks)} In \cite{Furman2006PremElliptical}, the concept of \emph{tail covariance premium} ($\mathrm{TCovP}$) was introduced, and it can be used as a capital allocation principle. Under $\mathrm{TCov}$ allocation, the capital $K_j$ allocated to the $j$-th risk is the $\theta$-level tail covariance premium \cite[Definition 4]{Furman2006PremElliptical}
\begin{equation}\label{eq:TCovPDef}
\mathrm{TCovP}_\theta(X_j|S) 
= \mathds{E}[X_j|S> \mathrm{V@R}_\theta(S)] + \beta \mathrm{Cov}(X_j,S|S> \mathrm{V@R}_\theta(S))
\end{equation}
for $1\le j\le M$, where $\beta\ge0$ is a constant. One has $\sum_{j=1}^M \mathrm{TCovP}_\theta(X_j|S) = \mathrm{TVP}_\theta(S)$, where $\mathrm{TVP}_\theta(S) = \mathds{E}[S|S> \mathrm{V@R}_\theta(S)] + \beta \mathrm{Var}(S|S> \mathrm{V@R}_\theta(S))$ is the \emph{tail variance premium} ($\mathrm{TVP}$) for the aggregate risk $S$ (see \cite[Definition 1]{Furman2006PremElliptical}), i.e. $\mathrm{TCov}$ allocation is additive when $\mathrm{TVP}_\theta(S)$ is used as a risk measure. If $\beta=0$, then $\mathrm{TCovP}_\theta(X_j|S)$ in (\ref{eq:TCovPDef}) reduces to the well known $\mathrm{TCE}$ allocation principle (which is also the $\mathrm{TV@R}$ allocation when the risks are continuous). Because $\mathds{E}[X_j|S>\mathrm{V@R}_\theta(S)]=\mathds{E}[X_j \mathds{1}\{S>\mathrm{V@R}_\theta(S)\}/(1-\theta)$ and
\[
\mathrm{Cov}(X_j,S|S> \mathrm{V@R}_\theta(S)) = \frac{\mathds{E}[X_jS \mathds{1}\{S> \mathrm{V@R}_\theta(S)\}]}{1-\theta} - \frac{\mathds{E}[X_j \mathds{1}\{S> \mathrm{V@R}_\theta(S)\}]\mathds{E}[S \mathds{1}\{S> \mathrm{V@R}_\theta(S)\}]}{(1-\theta)^2},
\]
the right-hand side of (\ref{eq:TCovPDef}) can be evaluated using Theorem \ref{th:JointTailMomentsXS} (with $y=\mathrm{V@R}_\theta(S)$) when $f\in\mathrm{MMEam}$.
\end{Example}
\begin{Example}\label{ex:TCPA} (\textbf{$\mathrm{TCPA}$ allocation for $\mathrm{MMEam}$ risks)} Noting that the expectation and covariance in (\ref{eq:TCovPDef}) are of different units, \cite{Wang2014TCovPA} proposed the \emph{tail covariance premium adjusted} ($\mathrm{TCPA}$) allocation rule which is additive for the $\theta$-level \emph{tail standard deviation premium} ($\mathrm{TSDP}$) discussed in \cite[Definition 2]{Furman2006PremElliptical}, namely $\mathrm{TSDP}_\theta(S) = \mathds{E}[S|S> \mathrm{V@R}_\theta(S)] + \beta \sqrt{\mathrm{Var}(S|S> \mathrm{V@R}_\theta(S))}$ for $\beta\ge0$. The $\mathrm{TCPA}$ allocation suggests that the allocated capital $K_j$ is
\[
\mathrm{TCPA}_\theta(X_j|S) = \mathds{E}[X_j|S> \mathrm{V@R}_\theta(S)] + \beta \frac{\mathrm{Cov}(X_j,S|S> \mathrm{V@R}_\theta(S))}{\sqrt{\mathrm{Var}(S|S> \mathrm{V@R}_\theta(S))}}
\]
for $1\le j\le M$. The components of $\mathrm{TCPA}_\theta(X_j|S)$ can be computed in the same manner as in Example \ref{ex:TCovRCA} when $f\in\mathrm{MMEam}$, and the additional term $\mathrm{Var}(S|S> \mathrm{V@R}_\theta(S))$ is simply $\mathrm{Var}(S|S>d) = \mathds{E}[(S-d)^2|S>d] - (\mathds{E}[S-d|S>d])^2$ evaluated at $d=\mathrm{V@R}_\theta(S)$ with the help of (\ref{eq:StopLossConditional}).
\end{Example}
\vspace{-.3in}

\subsection{Multiplicative background risk model and its capital allocation}\vspace{-.2in}
Assume that the vector of risks $\bm{X}=(X_1,\dots,X_M)$ is subject to some (extra) background risk $B>0$ which acts on all risks contained in $\bm{X}$ in a multiplicative manner \cite{franke2006multiplicative}. More specifically, suppose that the vector of interest, $\bm{X}^\dagger=(X_1^\dagger,\dots,X_M^\dagger)$, is of the form
\begin{align}\label{eq:BRM1}
(X_1^\dagger,\dots, X_M^\dagger) = \left(\frac{X_1}{B},\dots, \frac{X_M}{B}\right),
\end{align}
where $B$ is assumed to be independent of $\bm{X}$. The vector $\bm{X}^\dagger$ is then known as a \emph{multiplicative background risk model}, which can be used to describe a collection of risks that are affected by a common systemic component, such as regulatory constraints or some general economic conditions. Most of the works in the literature of multiplicative background risk models rely on the assumption that the risks $X_1,\dots,X_M$ are mutually independent, e.g. \cite{asimit2013evaluating, merz2014demand, Sarabia2016weightCA, su2017general, sarabia2018aggregation, furman2021multiplicative}. Some notable exceptions include \cite{zhu2012asymptotic} concerning asymptotic analysis, \cite{asimit2016background} who assumed $X_1,\dots,X_M$ have the same marginals with possibly different scaling factors, and \cite{cote2019dependence} who looked into the bivariate case. Our focus here is to analyze the background risk model (\ref{eq:BRM1}) under the assumption that $\bm{X}=(X_1,\dots,X_M)$ follows a density $f\in\mathrm{MMEam}$ as well as to provide formulas in relation to various capital allocation rules. This allows for models in which the individual risks exhibit interdependence before they are scaled by the systemic risk factor. We start by computing the marginal distributions of $(X_1^\dagger,\dots,X_M^\dagger)$.
\begin{theorem}\label{th:scaledXj} (\textbf{Individual risks in multiplicative background risk model under $\mathrm{MMEam}$}) Let $(X_1,\dots,X_M)$ follow the density $f\in\mathrm{MMEam}$ of the form (\ref{eq:densityMMEm1}) and let $B$ be an independent (strictly positive) risk with cumulative distribution function $G$ and Laplace transform $\mathcal{G}(z)=\int_0^\infty e^{-zr}\dd G(r)$. Under the background risk model $\bm{X}^\dagger= (X_1^\dagger,\dots,X_M^\dagger)$ defined via (\ref{eq:BRM1}), for $j\in\{1,\dots, M\}$ and $x\ge0$ the cumulative distribution function $F_{X_j^\dagger}$ of $X_j^\dagger$ is given by
\begin{equation}\label{eq:IPH1}
F_{X_j^\dagger}(x) = 1 - \sum_{\bm{i}\in\mathscr{S}} p_{\bm{i}}\bm{\alpha}_{i_j}\mathcal{G}(-\bm{T}_{i_j} x)  \bm{l}_{i_j},
\end{equation}
where for $k\in\{1,\dots,L\}$ we define
\begin{equation}\label{eq:LaplaceMatrix}
\mathcal{G}(-\bm{T}_kx)=\int_{0}^\infty e^{-(-\bm{T}_kx)r}\dd G(r).
\end{equation}
\end{theorem}\vspace{-.2in}
\begin{proof}
Conditioning on the value of $B$ followed by the use of (\ref{FXjDef}), we obtain
\begin{equation*}
F_{X_j^\dagger}(x) = 1 - \int_0^\infty \mathds{P}(X_j>xr)\dd G(r) = 1 - \int_0^\infty \left(\sum_{\bm{i}\in\mathscr{S}} p_{\bm{i}}\bm{\alpha}_{i_j} e^{\bm{T}_{i_j} xr} \bm{l}_{i_j} \right)\dd G(r),
\end{equation*}
from which (\ref{eq:IPH1}) follows under the definition (\ref{eq:LaplaceMatrix}).
\end{proof}\vspace{-.2in}

Although (\ref{eq:IPH1}) seemingly provides a closed-form expression for the cumulative distribution function of each risk $X_j^\dagger$, the Laplace transform $\mathcal{G}$ needs to be evaluated at a matrix argument. Nevertheless, computation of the right-hand side of (\ref{eq:LaplaceMatrix}) can be done by employing the canonical Jordan form as follows. First, recall that the Laplace transform $\mathcal{G}(z)=\int_0^\infty e^{-zr}\dd G(r)$ is holomorphic in $\mathds{C}_+=\{z\in\mathds{C}: \Re(z)>0\}$ \cite[Theorem 6.1]{doetsch2012introduction} and thus infinitely differentiable in $\mathds{C}_+$. For each fixed $k\in \{1,\dots,L\}$, define $\{\chi_{k,n}\}_{n=1}^{\ell_k}$ to be the eigenvalues of $-\bm{T}_{k}$, with $m_{k,n}$ the multiplicity of the eigenvalue $\chi_{k,n}$. The eigenvalues $\{\chi_{k,n}\}_{n=1}^{\ell_k}$ lie in $\mathds{C}_+$ (i.e. the set where $\mathcal{G}(z)$ is guaranteed to be holomorphic), and from \cite[Definition 1.2]{higham2008functions} it is known that
\begin{equation}\label{eq:Jordan1}
\mathcal{G}(-\bm{T}_{k}x)= \bm{Z}_k \mathcal{G}(\bm{J}_k x)\bm{Z}^{-1}_k,
\end{equation}
where the matrices $\bm{Z}_k$ and $\bm{J}_k=\mathrm{diag} (\bm{J}_{k,1},\ldots,\bm{J}_{k,\ell_k})$ define the canonical Jordan form of $-\bm{T}_k$ such that $-\bm{T}_k=\bm{Z}_k \bm{J}_k\bm{Z}^{-1}_k$, with the sub-blocks of $\bm{J}_k$ given by the $m_{k,n}$-dimensional matrices
\begin{equation*}
\bm{J}_{k,n}=
\begin{pmatrix}
\chi_{k,n} & 1 & &\\
&\chi_{k,n}& \ddots & \\
&&\ddots&1\\
&&&\chi_{k,n}
\end{pmatrix},
\end{equation*}
for $n=1,\ldots,\ell_k$. Then, $\mathcal{G}(\bm{J}_k x)$ in (\ref{eq:Jordan1}) can be represented as $\mathcal{G}(\bm{J}_k x)=\mathrm{diag}(\mathcal{G}(\bm{J}_{k,1} x),\ldots,\mathcal{G}(\bm{J}_{k,\ell_k} x))$, where
\begin{equation*}
\mathcal{G}(\bm{J}_{k,n} x) = \begin{pmatrix} \mathcal{G}(\chi_{k,n}x) &\frac{\mathcal{G}'(\chi_{k,n}x)}{1!}&\cdots&\frac{\mathcal{G}^{(m_{k,n}-1)}(\chi_{k,n}x)}{(m_{k,n}-1)!}\\&{\mathcal{G}(\chi_{k,n}x)}&\ddots&\vdots\\&&\ddots&\frac{\mathcal{G}'(\chi_{k,n}x)}{1!}\vspace*{1.5mm}\\
&&&{\mathcal{G}(\chi_{k,n}x)} \end{pmatrix}.
\end{equation*}
Alternatively, one can compute the right-hand side of (\ref{eq:LaplaceMatrix}) via the Cauchy integral formula 
\begin{equation}\label{eq:Cauchyintegral}
\mathcal{G}(-\bm{T}_k x) =\frac{1}{2\pi \mathrm{i}}\oint_\gamma \mathcal{G}(z) (z\bm{I}+\bm{T}_k x)^{-1}\dd z,\end{equation}
where $\oint$ denotes a line integral (in $\mathds{C}$), and $\gamma$ is any simple path contained in $\mathds{C}_+$ which encloses $\{\chi_{k,n}x\}_{n=1}^{\ell_k}$. That (\ref{eq:Jordan1}) and (\ref{eq:Cauchyintegral}) coincide follows by standard theory of holomorphic functions of matrices (see e.g. \cite[Chapter 1]{higham2008functions}). In general, the form (\ref{eq:Cauchyintegral}) is useful for algebraic manipulations whereas (\ref{eq:Jordan1}) allows for easy numerical implementation.



Now let us define the aggregate multiplicative background risk $S^\dagger = X_1^\dagger+\cdots + X_M^\dagger= S/B$. Since $S$ follows the $\mathrm{MEam}$ density $f_S$ in (\ref{eq:fSresult}), employing similar steps to those in the proof of Theorem \ref{th:scaledXj} gives the cumulative distribution function of $S^\dagger$, namely
\begin{equation*}
F_{S^\dagger}(x)=1 -  \sum_{\bm{i}\in\mathscr{S}} p_{\bm{i}} \bm{\alpha}_{\bm{i}}\mathcal{G}(-\bm{T}_{\bm{i}} x) \bm{l}_{\bm{i}},
\end{equation*}
for $x\ge0$, where $\bm{l}_{\bm{i}}=(-\bm{T}_{\bm{i}})^{-1}\bm{t}_{\bm{i}}$. The above equation can be used to compute $\mathrm{V@R}_\theta(S^\dagger)$ via $F_{S^\dagger}(\mathrm{V@R}_\theta(S^\dagger))=\theta$, which is needed in capital allocation rules in relation to those discussed in Section \ref{subsec:riskcapital1}. For example, utilizing the independence assumption between $B$ and $\bm{X}$, the covariance-based allocation (see Example \ref{ex:CapitalCov}) for the multiplicative background risk model $\bm{X}^\dagger=(X_1^\dagger,\dots,X_M^\dagger)$ is given by
\begin{equation*}
C_\theta(X_j^\dagger,S^\dagger) 
=  \mathds{E}[X_j]\mathds{E}[B^{-1}] + \frac{\mathds{E}[X_jS]\mathds{E}[B^{-2}]-\mathds{E}[X_j]\mathds{E}[S](\mathds{E}[B^{-1}])^{-2}}{\mathds{E}[S^2]\mathds{E}[B^{-2}]-(\mathds{E}[S]\mathds{E}[B^{-1}])^{-2}} (\mathrm{TV@R}_\theta(S^\dagger)-\mathds{E}[S]\mathds{E}[B^{-1}])
\end{equation*}
provided that $\mathds{E}[B^{-2}]$ is finite. While the calculations of $\mathds{E}[X_j]$, $\mathds{E}[S]$, $\mathds{E}[S^2]$ and $\mathds{E}[X_jS]$ have already been discussed in Example \ref{ex:CapitalCov}, we additionally require $\mathrm{TV@R}_\theta(S^\dagger)$ in the above formula. Since $S^\dagger$ is a continuous random variable, one has $\mathrm{TV@R}_\theta(S^\dagger) = \mathrm{TCE}_\theta(S^\dagger) = \mathds{E}[S^\dagger|S^\dagger>\mathrm{V@R}_\theta(S^\dagger)]$. 
If one is interested in $\mathrm{TCov}$ and $\mathrm{TCPA}$ allocations corresponding to Examples \ref{ex:TCovRCA} and \ref{ex:TCPA} respectively, then in the model $\bm{X}^\dagger$ these are given by
\begin{equation*}
\mathrm{TCovP}_\theta(X_j^\dagger|S^\dagger) = \mathds{E}[X_j^\dagger|S^\dagger> \mathrm{V@R}_\theta(S^\dagger)] + \beta \mathrm{Cov}(X_j^\dagger,S^\dagger|S^\dagger> \mathrm{V@R}_\theta(S^\dagger)),
\end{equation*}
and
\begin{equation*}
\mathrm{TCPA}_\theta(X_j^\dagger|S^\dagger) = \mathds{E}[X_j^\dagger|S^\dagger> \mathrm{V@R}_\theta(S^\dagger)] + \beta \frac{\mathrm{Cov}(X_j^\dagger,S^\dagger|S^\dagger> \mathrm{V@R}_\theta(S^\dagger))}{\sqrt{\mathrm{Var}(S^\dagger|S^\dagger> \mathrm{V@R}_\theta(S^\dagger))}}.
\end{equation*}
It is instructive to note that all the afore-mentioned three allocation rules for $\bm{X}^\dagger$ can be fully characterized if we can compute the (cross) moments in the form of $\mathds{E}[(X^\dagger_j)^k(S^\dagger)^h\mathds{1}\{S^\dagger>y\}]$ concerning the individual risk $X_j^\dagger$ and the aggregate risk $S^\dagger$. Such a quantity is provided in Theorem \ref{th:JointTailMomentsXSMBR} below, which is analogous to Theorem \ref{th:JointTailMomentsXS}.
\begin{theorem}\label{th:JointTailMomentsXSMBR}  (\textbf{Cross moments of individual and total multiplicative background risks under $\mathrm{MMEam}$}) Let $(X_1,\dots,X_M)$ follow the density $f\in\mathrm{MMEam}$ of the form (\ref{eq:densityMMEm1}) and let $B$ be an independent (strictly positive) risk with cumulative distribution function $G$ and Laplace transform $\mathcal{G}(z)=\int_0^\infty e^{-zr}\dd G(r)$. Fix $M\ge 2$, $j\in\{1,\dots, M\}$ and consider the cross moments of $X_j^\dagger$ and $S^\dagger = \sum_{i=1}^MX_i^\dagger$ on the set $\{S^\dagger>y\}$ in the background risk model $\bm{X}^\dagger= (X_1^\dagger,\dots,X_M^\dagger)$ defined via (\ref{eq:BRM1}). For $h\in\mathds{N}_0$ and $y\ge 0$, we have
\begin{equation}\label{eq:MomentsXSk0MBR}
\mathds{E}[(S^\dagger)^h\mathds{1}\{S^\dagger>y\}] = \sum_{\bm{i}\in\mathscr{S}} p_{\bm{i}} \bm{\alpha}_{\bm{i}} \left(\sum_{\ell=0}^h (-\bm{T}_{\bm{i}})^{-(h-\ell+1)} \mathcal{G}_{\ell-h}(-\bm{T}_{\bm{i}}y) \frac{h!}{\ell!} y^\ell \right) \bm{t}_{\bm{i}}
\end{equation}
if $\mathds{E}[B^{-h}]<\infty$, where
\begin{equation}\label{eq:higherLaplaceMatrix1}
\mathcal{G}_{\ell-h}(-\bm{T}_{\bm{i}}y) = \int_0^\infty r^{\ell-h}e^{-(-\bm{T}_{\bm{i}}y)r}\dd G(r),
\end{equation}
for $\ell\in\{0,\dots, h\}$. For $k\in\mathds{N}_+$, $h\in\mathds{N}_0$ and $y\ge 0$, we have
\begin{equation}\label{eq:capitalaux1MBR}
\mathds{E}[(X^\dagger_j)^k(S^\dagger)^h\mathds{1}\{S^\dagger>y\}] = \sum_{\bm{i}\in\mathscr{S}} p_{\bm{i}}  k! \left(\bm{\alpha}_{i_j}^{[k]}, \bm{0}\right)\left\{\sum_{\ell=0}^h \left(-\bm{A}^{\{k,\bm{i},j\}}\right)^{-(h-\ell+1)} \mathcal{G}_{\ell-k-h}(-\bm{A}^{\{k,\bm{i},j\}}y)\frac{h!}{\ell!} y^\ell \right\}\begin{pmatrix}\bm{0}\\ \bm{t}_{(\bm{i},j)}\end{pmatrix}
\end{equation}
if $\mathds{E}[B^{-k-h}]<\infty$, where 
\begin{equation}\label{eq:higherLaplaceMatrix2}
\mathcal{G}_{\ell-k-h}(-\bm{A}^{\{k,\bm{i},j\}}y) = \int_0^\infty r^{\ell-k-h}e^{-(-\bm{A}^{\{k,\bm{i},j\}}y)r}\dd G(r),
\end{equation}
for $\ell\in\{0,\dots, h\}$.
\end{theorem}\vspace{-.2in}
\begin{proof}
Conditioning on the value of $B$ yields
\begin{equation*}
\mathds{E}[(S^\dagger)^h\mathds{1}\{S^\dagger>y\}] = \int_0^\infty r^{-h}\mathds{E}[S^h\mathds{1}\{S>yr\}] \dd G(r)
\end{equation*}
and
\begin{equation*}
\mathds{E}[(X^\dagger_j)^k(S^\dagger)^h\mathds{1}\{S^\dagger>y\}] = \int_0^\infty r^{-k-h}\mathds{E}[X^k_jS^h\mathds{1}\{S>yr\}] \dd G(r).
\end{equation*}
The results (\ref{eq:MomentsXSk0MBR}) and (\ref{eq:capitalaux1MBR}) follow by plugging (\ref{eq:MomentsXSk0}) and (\ref{eq:capitalaux1}) into the above expressions. Note that the condition $\mathds{E}[B^{-h}]<\infty$ guarantees that the right-hand side of (\ref{eq:higherLaplaceMatrix1}) is finite for all $\ell\in \{0,\dots,h\}$, and similar comments apply to (\ref{eq:higherLaplaceMatrix2}).
\end{proof}\vspace{-.2in}

As long as we can evaluate the matrices $\mathcal{G}_{\ell-h}(-\bm{T}_{\bm{i}}y)$ and $\mathcal{G}_{\ell-k-h}(-\bm{A}^{\{k,\bm{i},j\}}y)$, Theorem \ref{th:JointTailMomentsXSMBR} provides closed-form formulas for the cross moments of $X_j^\dagger$ and $S^\dagger$ on the event $\{S^\dagger > y\}$. Fortunately, the right-hand side of (\ref{eq:higherLaplaceMatrix1}) and (\ref{eq:higherLaplaceMatrix2}) can be computed via the method of canonical Jordan form as in (\ref{eq:Jordan1}) or the Cauchy integral formula (\ref{eq:Cauchyintegral}) with the obvious modifications.
\begin{Remark} To employ the canonical Jordan form or the Cauchy integral formula to compute the function of a matrix, we need to first verify that the function in question is holomorphic in $\mathds{C}_+$ (where the eigenvalues of $-\bm{T}_{\bm{i}}$ and $-\bm{A}^{\{k,\bm{i},j\}}$ lie). If $\mathds{E}[B^{-\ell}]=\int_0^\infty r^\ell \dd G(r)<\infty$ (for a given $\ell < 0$), then $\mathcal{G}_{\ell}(z)=\int_0^\infty r^\ell e^{-zr} \dd G(r)$ converges in the region $z\in\mathds{C}_+$. Because $\mathcal{G}_{\ell}$ can be regarded as the Laplace transform in relation to $r^\ell \dd G(r)$, it must be analytic in $\mathds{C}_+$. 
\end{Remark}\vspace{-.3in}

\section{A note on calibration with data and future research}\label{sec:calibration}\vspace{-.2in}
In this section, we briefly address the problem of performing statistical inference of $f\in\mathrm{MMEam}$ given complete data. The method outlined here employs existing methods to fit phase-type and $\mathrm{ME}$ distributions to univariate data, as well as elements of the theory of copulas. In short, our methodology consists of the estimation of the $\mathrm{ME}$ marginals followed by fitting of a dependence structure. By choosing a particular but robust dependence structure in the second step via a Bernstein copula, it is guaranteed that the fitted density is an element of $\mathrm{MMEam}$. At the end we will conclude by discussing alternative directions of research regarding statistical inference for $\mathrm{MMEam}$ densities.

Suppose that we have $N$ observed realizations, $\mathcal{D}=\{(x_{k,1}, \dots, x_{k,M})\}_{k=1}^N$, of a positive $M$-variate random vector $\bm{X}=(X_1,\dots,X_M)$ for which we want to produce a statistical fitting $\widehat{F}$ of its multivariate cumulative distribution function $F$. Through Sklar's theorem \cite[Theorem 2.3.3]{nelsen2007introduction}, the theory of copulas allows us to calibrate $F$ in two steps:\vspace{-.2in}
\begin{enumerate}
  \item For each $j\in\{1,\dots, M\}$, fit a distribution function $\widehat{F}_{X_j}$ to the marginal data $\{ x_{k,j}\}_{k=1}^N$.
  \item Using the dependency structure of the data, construct a copula function $\widehat{C}:[0,1]^M\rightarrow [0,1]$ \cite[Definition 2.2.2]{nelsen2007introduction}, and let the fitted distribution $\widehat{F}$ of $F$ be defined by
\begin{equation}\label{eq:Fcopula1}
\widehat{F}(x_1,\dots,x_m) = \widehat{C}(\widehat{F}_{X_1}(x_1),\dots, \widehat{F}_{X_M}(x_M)).
\end{equation}
\end{enumerate}\vspace{-.2in}

In what follows, we expand on each of the steps applied to the $\mathrm{MMEam}$ framework.

\noindent\textbf{Fitting the marginals.} Fix $j\in\{1,\dots,M\}$ and let $\mathcal{D}_j=\{ x_{k,j}\}_{k=1}^N$. The problem of fitting an $\mathrm{ME}$ distribution $\widehat{F}_{X_j}$ to the data $\mathcal{D}_j$ has mainly been studied in the literature for the phase-type subclass only. Most of the existing fitting methods for phase-type distributions fall under the umbrella of maximum likelihood estimation, moment matching, or Bayesian inference (see \cite{okamura2016fitting} for a comprehensive analysis of phase-type fitting methods). In general, there is no method that outperforms the others, as fitting with phase-type distributions is (in most cases) model specific. But the Expectation-Maximization method proposed in \cite{asmussen1996fitting} is possibly the most adopted one due to its flexibility. We note that an $\mathrm{ME}$-specific maximum likelihood method considered in \cite{fackrell2005fitting} may potentially provide $\mathrm{ME}$ fits with lower dimensions than those resulting from phase-type specific methods. However, it requires the minimization of a convex function under uncountably many constraints, which results in high computational costs and limits its practical use. Using either of the outlined methods, let $\widehat{f}_{X_j}(x)=\widehat{\bm{\alpha}}_{X_j}e^{\widehat{\bm{T}}_{X_j} x}\widehat{\bm{t}}_{X_j}$ be the $\mathrm{ME}$ (possibly phase-type) density function which fits the data $\mathcal{D}_j$, i.e. $\widehat{f}_{X_j}\in\mathrm{ME}$ with parameters $(\widehat{\bm{\alpha}}_{X_j}, \widehat{\bm{T}}_{X_j},\widehat{\bm{t}}_{X_j})$. For later use we also define $\widehat{\bm{l}}_{X_j}=(-\widehat{\bm{T}}_{X_j})^{-1}\widehat{\bm{t}}_{X_j}$ so that $\widehat{F}_{X_j}(x)=1-\widehat{\bm{\alpha}}_{X_j}e^{\widehat{\bm{T}}_{X_j} x}\widehat{\bm{l}}_{X_j}$.

\noindent\textbf{Constructing the dependency structure.} Once the marginal $\widehat{f}_j\in\mathrm{ME}$ has been estimated for each $j\in\{1,\dots,M\}$, the next task is to construct an appropriate copula $\widehat{C}$ which can capture the dependency among the marginals through (\ref{eq:Fcopula1}). The main challenge here is to choose $\widehat{C}$ in such a way that the corresponding fitted joint density $\widehat{f}$ is an $\mathrm{MMEam}$ density, and thus the rich theory presented throughout this paper is applicable. We propose employing the Bernstein copula $C:[0,1]^M\rightarrow [0,1]$ of order $A\in \mathds{N}_+$ taking the form \cite{sancetta2004bernstein}
\begin{equation}\label{eq:BernsteinC1}
C(u_1,\dots, u_M) = \sum_{(h_1,\dots,h_M)\in \{0,\dots,A\}^M} \zeta\left(\frac{h_1}{A},\dots,\frac{h_M}{A}\right) \prod_{j=1}^M{A\choose h_j} u_j^{h_j}(1-u_j)^{A-h_j},
\end{equation}
where $\{\zeta(h_1/A,\dots,h_M/A)\}_{(h_1,\dots,h_M)\in \{0,\dots,A\}^M}$ is a collection of real numbers satisfying certain conditions to ensure that $C$ is a copula (see \cite[Theorem 1]{sancetta2004bernstein}). It can be shown \cite[Lemma 1]{sancetta2004bernstein} that the Bernstein copula (with varying $A$) is dense within the set of $M$-dimensional copulas, meaning that any dependence structure can be approximated arbitrarily well by (\ref{eq:BernsteinC1}). If the dependence structure is known to exactly follow the above Bernstein copula, then with the fitted marginals $\{\widehat{F}_{X_j}\}_{j=1}^M$ the joint cumulative distribution function of $(X_1,\dots,X_M)$ can be estimated as $C(\widehat{F}_{X_1}(x_1),\dots, \widehat{F}_{X_M}(x_M))$. In order to mirror the dependence actually present in the data, we propose to use the empirical mixing weights $\{\widehat{\zeta}(h_1/A,\dots,h_M/A)\}_{(h_1,\dots,h_M)\in \{0,\dots,A\}^M}$ \cite[Definition 1 and Section 4.1]{sancetta2004bernstein} of the form
\begin{equation}\label{eq:qhat1}
\widehat{\zeta}\left(\frac{h_1}{A},\dots,\frac{h_M}{A}\right)= \frac{1}{N} \sum_{k=1}^N \mathds{1}\left\{\bigcap_{j=1}^M \left\{\widehat{F}_{X_j}(x_{k,j})\le \frac{h_j}{A}\right\}\right\},
\end{equation}
and the copula (\ref{eq:BernsteinC1}) with $\widehat{\zeta}$ in place of $\zeta$ will be denoted by $\widehat{C}$. The copula density corresponding to $\widehat{C}$ is \cite[Section 2.2]{sancetta2004bernstein}
\begin{align*}
\widehat{c}(u_1,\dots, u_M)  =&~A^M \sum_{(h_1,\dots,h_M)\in \{0,\dots,A-1\}^M} \widehat{\phi}_{(h_1,\dots,h_M)} \prod_{j=1}^M{A-1\choose h_j} u_j^{h_j}(1-u_j)^{A-1-h_j}\\
=&~\sum_{(h_1,\dots,h_M)\in \{0,\dots,A-1\}^M} \widehat{\phi}_{(h_1,\dots,h_M)} \prod_{j=1}^M \frac{A!}{(A-1-h_j)!h_j!} u_j^{h_j}(1-u_j)^{A-1-h_j},
\end{align*}
where
\begin{equation*}
\widehat{\phi}_{(h_1,\dots,h_M)} =  \sum_{(\ell_1,\dots,\ell_M)\in \{0,1\}^M}(-1)^{M+\ell_1+\ldots+\ell_M} \widehat{\zeta}\left(\frac{h_1+\ell_1}{A},\dots,\frac{h_M+\ell_M}{A}\right)
\end{equation*}
is expressed in terms of (\ref{eq:qhat1}). The fitted joint density of $(X_1,\dots,X_M)$ is thus
\begin{align*}
&\widehat{f}(x_1,\dots,x_m) = \widehat{c}(\widehat{F}_{X_1}(x_1),\dots, \widehat{F}_{X_M}(x_M)) \prod_{j=1}^M \widehat{f}_{X_j}(x_j)\\
&\quad = \sum_{(h_1,\dots,h_M)\in \{0,\dots,A-1\}^M} \widehat{\phi}_{(h_1,\dots,h_M)} \prod_{j=1}^M \frac{A!}{(A-1-h_j)!h_j!} \left(\widehat{F}_{X_j}(x_j)\right)^{h_j} \left(1-\widehat{F}_{X_j}(x_j)\right)^{A-1-h_j} \widehat{f}_{X_j}(x_j)\\
&\quad = \sum_{(h_1,\dots,h_M)\in \{0,\dots,A-1\}^M} \widehat{\phi}_{(h_1,\dots,h_M)} \prod_{j=1}^M \widehat{f}_{X_j,h_j+1;A}(x_j),
\end{align*}
where $\widehat{f}_{X_j,h_j+1;A}$ is the density of the $(h_j+1)$-th order statistic of $A$ independent variables with common density $\widehat{f}_{X_j}$. By rewriting the above equation as
\begin{equation*}
\widehat{f}(x_1,\dots,x_m) =  \sum_{(h_1,\dots,h_M)\in \{1,\dots,A\}^M} \widehat{\phi}_{(h_1-1,\dots,h_M-1)} \prod_{j=1}^M \widehat{f}_{X_j,h_j;A}(x_j)
\end{equation*}
and recalling that each $\widehat{f}_{X_j,h_j;A}$ is an $\mathrm{ME}$ density (see Proposition \ref{prop:order1} or Remark \ref{rem:orderstats2}), we can conclude that the above fitted joint density $\widehat{f}$ is in the form of an $\mathrm{MMEam}$ density as in (\ref{eq:densityMMEm1}). Interested readers are also referred to e.g. \cite{baker2008order} for further connection between Bernstein copula and order statistics.

Let us conclude by discussing a few points about our calibration scheme. The benefit of our method is that practitioners are able to choose among existing univariate fitting schemes for the $\mathrm{ME}$ class or its subclasses (such as phase-type), with the choice possibly depending on the model and the data in hand. Modelling dependence through the Bernstein copula is a simple (and to some extent naive) solution, especially when using the empirical mixing weights. On the other hand, to get a reasonably good approximation of the dependence structure of the data, we likely need to choose a large $A$, meaning that the $\mathrm{ME}$ density $\widehat{f}_{X_j,h_j;A}$ will be high dimensional. To improve this, one possible direction of research is to investigate more efficient mixing weights for the Bernstein copula like those proposed in \cite{dou2016algorithms} obtained through an Expectation-Maximization algorithm. Furthermore, as pointed out in \cite{mikosch2006copulas}, performing statistical fitting in a two-step fashion like the one proposed here creates two different types of approximation errors, which might in turn cause a miscalibration of the data. Recently, statistical inference for different classes of multivariate phase-type distributions has been performed by \cite{albrecher2020fitting} in a single step (i.e. magnitude and dependence are calibrated at the same time) by extending the work of \cite{asmussen1996fitting}. A possible direction of research is to adapt their work to produce maximum likelihod estimation methods for the  $\mathrm{MMEam}$ class.\vspace{-.2in}

\section*{Acknowledgements} \vspace{-.2in}
OP acknowledges the funding of the Australian Research Council's Discovery Project DP180103106. EC and JKW acknowledge the support from the Australian Research Council's Discovery Project DP200100615.

\end{document}